# Uncovering the nucleus candidate for NGC 253

G. I. Günthardt

Observatorio Astronómico, Universidad Nacional de Córdoba, Argentina.

gunth@oac.uncor.edu

M. P. Agüero

Observatorio Astronómico, Universidad Nacional de Córdoba, and CONICET, Argentina.

mpaguero@oac.uncor.edu

J. A. Camperi

Observatorio Astronómico, Universidad Nacional de Córdoba, Argentina.

camperi@oac.uncor.edu

R.J. Díaz

Gemini Observatory, AURA, USA.

ICATE, CONICET, Argentina.

rdiaz@gemini.edu

P.L. Gomez

Gemini Observatory, AURA, USA.

pgomez@gemini.edu

G. Bosch

Instituto de Astrofísica de La Plata (CONICET-UNLP), Argentina.

guille@fcaglp.unlp.edu.ar

and

M. Schirmer

Gemini Observatory, AURA, USA.

Argelander-Institut fur Astronomie, Universitat Bonn, Germany.

mschirmer@gemini.edu

## ABSTRACT

NGC 253 is the nearest spiral galaxy with a nuclear starburst that becomes the best candidate for studying the relationship between starburst and active galactic nucleus activity. However, this central region is veiled by large amounts of dust, and it has been so far unclear which is the true dynamical nucleus to the point that there is no strong evidence that the galaxy harbors a supermassive black hole co-evolving with the starburst as was supposed earlier. Near-infrared (NIR) spectroscopy, especially the NIR emission line analysis, could be advantageous in order to shedding light on the true nucleus identity. Using Flamingos-2 at Gemini South we have taken deep K-band spectra along the major axis of the central structure and through the brightest infrared source. In this work, we present evidence showing that the brightest NIR and mid-infrared source in the central region, already known as radio source TH7 and so far considered just a large stellar supercluster, in fact presents various symptoms of a genuine galactic nucleus. Therefore, it should be considered a valid nucleus candidate. Mentioning some distinctive aspects, it is the most massive compact infrared object in the central region, located at 2.0" of the symmetry center of the galactic bar, as measured in the K-band emission. Moreover, our data indicate that this object is surrounded by a large circumnuclear stellar disk and it is also located at the rotation center of the large molecular gas disk of NGC 253. Furthermore, a kinematic residual appears in the $H_2$ rotation curve with a sinusoidal shape consistent with an outflow centered in the candidate nucleus position. The maximum outflow velocity is located about 14 pc from TH7, which is consistent with the radius of a shell detected around the nucleus candidate, observed at 18.3 μm (Qa) and 12.8 μm ([NeII]) with T-ReCS. Also, the Brγ emission line profile shows a pronounced blueshift and this emission line also has the highest equivalent width at this position. All this evidence points to TH7 as the best candidate for the galactic nucleus of NGC 253.

# 1. Introduction

Recent advances in instrumental capabilities have triggered some hypotheses and debate about the connection between circumnuclear starbursts and active galactic nuclei (Cid-Fernandes et al. 2001; Chen et al. 2009; Gonzalez Delgado et al. 2009; Rodríguez-Ardila et al. 2009; Rafanelli et al. 2011; Davies et al. 2014). Theses speculations are due mostly to the fact that an increasing number of the neighboring galaxies, as revealed by new observations, contain starbursts and active galactic nuclei (AGNs) in close proximity (see Levenson et al. 2001 and references therein). It remains unclear whether or not proximity implies a physical connection, but a mechanism for forming a supermassive black hole (and subsequent AGN) certainly could involve a circumnuclear starburst, given that it can process as much as $\sim 10^{10}$ $M_\odot$ of material in $10^7 - 10^8$ years (Norman & Scoville 1988). Nevertheless, evidence for such a scenario has been scarce until recently. Starburst galaxies, which show a star formation rate occasionally in excess of $\sim 100$ $M_\odot$ yr$^{-1}$ (Kennicutt 1998), are very useful for studying earlier eras of star formation in the Universe. These galaxies are consuming their available gas at a rate that is not sustainable over a Hubble time. In local starburst galaxies such as NGC 253, star forming regions are resolved into dense super star clusters (SSCs), which are the most massive example ($\sim 10^6$ $M_\odot$) of clustered star formation. These SSCs, with core stellar densities in excess of $10^4$ pc$^{-3}$ (Johnson 2005), are the outcome of star formation in an extreme environment. The most massive SSCs can eventually evolve into the most massive object in the host galaxy and provide a link in the coevolution of SMBH and the stellar systems which harbor them (e.g. Díaz et al. 2006; Rodrigues et al. 2009). Recent developments in the numerical simulations of barred galaxies have shown that the interplay between the galactic bar and a supermassive black hole yield to the decoupling of the central cluster from the large scale structures, via interactions with the black hole, and the presence of a compact object at the center of the galaxy will play a dominant role on the how and where the gas exchanges angular momentum and form stars (e.g. Emsellem et al. 2015 and references therein).

Local starburst galaxies can be observed at the scale necessary to resolve the basic elements of this scenario. At a distance of 3.9 ± 0.37 Mpc (Karachentsev et al. 2003) where 1" ~ 17 pc, NGC 253 is a close example of a rather typical starburst galaxy in the Sculptor group (α ~ 00$^h$ 47$^m$ , δ ~ −25°17'). The starburst nature of NGC 253 is believed to be triggered from the presence of a 6 kpc bar that funnels gas into the nucleus (Engelbracht et al. 1998). The presence of compact radio sources corresponding to H II regions and [Fe II] sources tracing supernova remnants (SNRs) have been observed in the inner starburst disk (radius 15" − 20" ; 255–340 pc) of NGC 253 (Ulvestad & Antonucci 1997; Alonso-Herrero et al. 2003). Consistently, Lenc & Tingay (2006) report a high supernova rate of > 0.14 yr$^{-1}$. Multiple discrete sources have been observed at several wavelengths, including approximately 60 compact radio sources. This includes the radio peak of NGC 253 at 2 cm, TH2, which was thought to be the kinematic center of the galaxy (Turner & Ho 1985). So far, detailed studies of the nuclear environment have failed to identify a discrete source associated with TH2 that could be called the galaxy nucleus (Fernández-Ontiveros et al. 2009). On the other hand, Müller-Sánchez et al. (2010) find a stellar kinematic center (SKC) located between the strongest compact radio source (TH2) and the IR peak, that is, 2.6" NE from the IR peak.

Here, we investigate the properties of the IR brightest objects in the circumnuclear region of NGC

253, including the neighborhood of the former nucleus candidate TH2 and the object of study of the present work, the SSC coincident with the near-infrared (NIR) and mid-infrared (MIR) emission peak (Kornei & McCrady 2009). For this purpose, we have performed observations with the new instrumental facility of Gemini South, Flamingos-2 (NIR wide field imager and multi-object spectrometer, Eikenberry et al. 2008; Gomez et al. 2012). With this instrument we have obtained NIR images and NIR longslit Ks-band spectra along the major axis of NGC 253. Additionally, we analyzed MIR images from T-ReCS (Gemini South) (Telesco et al. 1998).

This paper is structured as follows. In Section 2, we describe the Gemini Flamingos-2 and T-ReCS observations. The spectroscopy and image data reduction procedure is mentioned. Also, the Flamingos-2 performance is discussed. In Section 3, the analysis of the circumnuclear and inner structures is carried out (Section 3.1 and 3.2 respectively) as seen in NIR and MIR imaging. Similarly, the NIR spectroscopic and kinematic properties are studied. The emission line and radial profiles are analyzed (Section 3.3). An NIR diagnostic diagram of some nuclear structures is presented in Section 3.4 and the study of EW(Brγ) and FWHM(Brγ) of those zones is examined in Section 3.5. The $H_2$ molecular gas rotation curve is constructed and the kinematical perturbations are modeled in Section 3.6. In Section 4, the discussion of the infrared brightness peak core as a nucleus candidate is developed. Finally, in Section 5 the conclusions are submitted. A summary of the main structures is provided in Figure 19 and Table 4.

## 2. Observations

NIR observations have been performed with Flamingos-2 (hereafter F2) at the 8.1 m Gemini South Telescope during a commissioning run in 2013 June 24 and broadband Y, J, H, Ks images were obtained before the spectroscopic acquisition with a total exposure time of 18 s per filter (Table 1). Sky frames were taken 10' away from the galaxy main body and were also used to construct the flat field images. The median seeing at the Ks images is 0.5". The images were flux calibrated using the zero point calibrations for the photometric MKO system obtained using 88 independent data points from standard star observations made during the three commissioning runs of 2013 May through July. These zero points were obtained using the software package THELI (Schirmer 2013):

$Y = 25.12 - 0.01*k + 0.50*(Y-J) \pm 0.03$
$J = 25.21 - 0.02*k + 0.87*(J-H) \pm 0.05$
$H = 25.42 - 0.01*k + 0.73*(J-H) \pm 0.05$
$Ks = 24.64 - 0.05*k - 0.27*(H-Ks) \pm 0.06$

where k is the extinction coefficient.

We obtained spectra using the R3000 grism with a Ks filter, covering a wavelength range from 1.95 to 2.35 μm. The effective on-source total exposure time was 2400 s. Sky spectra were obtained 7 arcmin away from the galaxy main body. The 3 pixel slit provided a peak resolution R~1600 at about 2.15 μm, dropping to about R~800 at the spectral ends. This change in resolution is a result of the usual grism variable resolution. The F2 initial performance is reported in Díaz et al (2013). During all the F2 observations in 2013 the spatial resolution was a function of radius: with good seeing the nominal resolution of 0.35" was achieved with no radial dependence until 1.5′, and afterward it consistently dropped across the whole field as function of radius (Díaz et al. 2013). This problem was corrected in 2014 and at present, the spatial resolution is uniform across the whole 6′ field of view (FOV). The study reported here focuses on the central 2′ of the slit and is not affected by the F2 camera performance. All the spectroscopic data have been pre-reduced using the F2 Gemini pipeline and the reduction was completed following standard reduction techniques with IRAF.

Figure 1 shows the central 20" of the nuclear region of NGC 253 in a Paschen α filter (F187N) obtained from archive data from the Hubble Space Telescope (HST),. The relevant portion of the slit position is superimposed in the image.

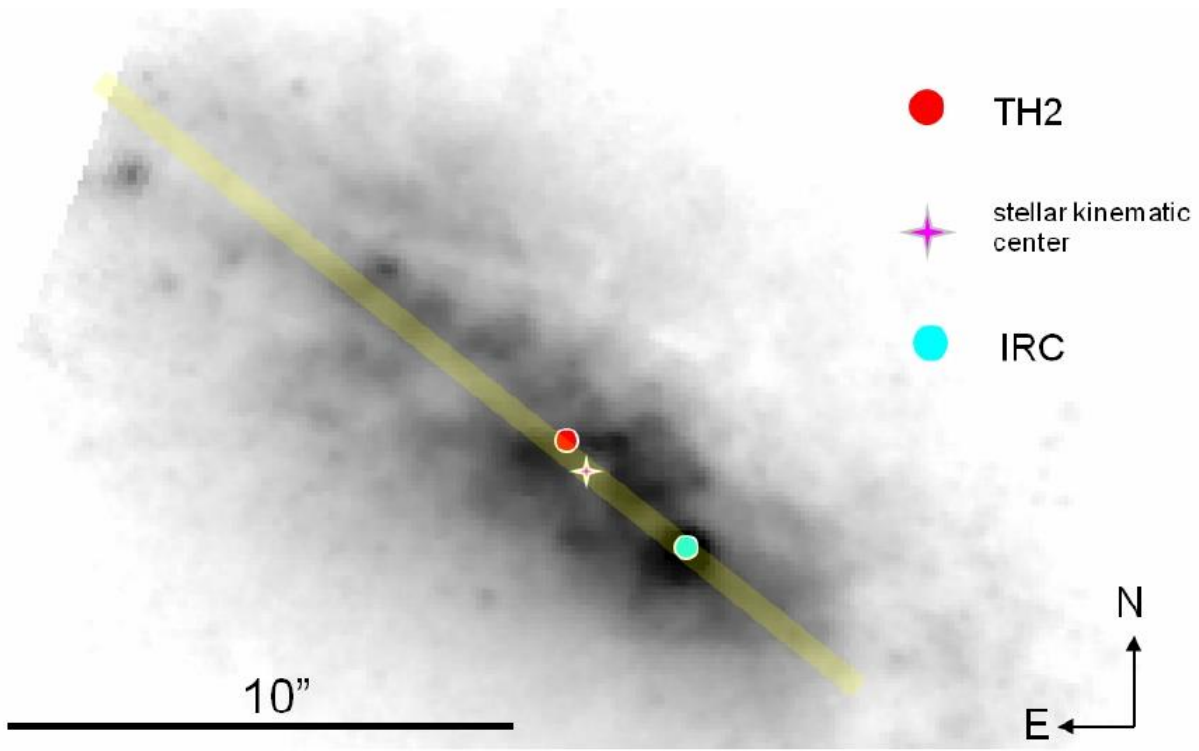

*Figure 1.* Pα (F187N filter) HST image of the central ~ 20" of the nuclear region of NGC 253. The position and width of the slit are marked. The positions of TH2, IRC (TH7), and the stellar kinematic center (Müller-Sánchez et al. 2010) are also included. The intensity display of the image is in logarithmic scale.

MIR images (see Table 1) were obtained with T-ReCS at Gemini in 2011 August 11 through the following filters: Qa (18.3 μm), Si-2 (8.7 μm), [NeII] (12.8 μm), [NeII] continuum (13.1 μm) totaling two hours of observation including overheads. These images were obtained as part of a T-ReCS legacy program lead by B. Rodgers (GS-2011B-Q-84) and were processed with the THELI software package (Erben et al. 2005; Schirmer 2013). The scale of the T-ReCS detector is of 0.09"/pixel.

All the circumnuclear spatial positions reported in the spectra and images are within the central arcminute of F2 FOV. Using 2MASS stars, we determined that NGC 253 observations have a scale of (0.1798 ± 0.0015)"/pixel in the central 2 arcmin of the FOV. This scale value and the stability of the astrometric distortion were verified using M17 images taken on the same night. Therefore all the relative positions discussed in the paper are actually as accurate as the centroid's determination, which was better than 1 pixel (0.18") in all cases. The absolute coordinates of T-ReCS knots derived from the images' WCS have an uncertainty of 0.3" (rms) when compared with the Q-band images published by Fernández-Ontiveros et al. (2009), which is consistent with the astrometric precision reported at the Gemini instrument web-pages. Centroid matching between matched NIR and MIR purely pointlike sources is consistent within 0.1″ for knots A1, A6, A8, and A9 (the notation for the knots' denominations is detailed in Section 3.1), indicating that astrometric distortions can be neglected within the small FOV studied in this paper.

Table 1
Imaging Configuration

| Instrument | Filter | $\lambda_0$ ($\mu$m) | $\Delta\lambda$ (50% trans., nm) | FWHM (") | Total Exp. Time (s) |
|---|---|---|---|---|---|
| F2 | Y-G0811 | 1.020 | 99 | 0.7 | 18 |
| F2 | J-G0802 | 1.255 | 158 | 0.6 | 18 |
| F2 | H-G0803 | 1.631 | 289 | 0.6 | 18 |
| F2 | Ks-G0804 | 2.157 | 329 | 0.4 | 18 |
| T-ReCS | Si-2 | 8.7 | 780 | 0.4 | 521.3 |
| T-ReCS | [Ne II] | 12.8 | 230 | 0.5 | 551.6 |
| T-ReCS | [Ne II] continuum | 13.1 | 220 | 0.5 | 551.6 |
| T-ReCS | Qa | 18.3 | 1510 | 0.6 | 579.2 |

Notes. Col. (1): instrument, col. (2): filter denomination, col. (3): filter central wavelength, col. (4): 50% cut-on/off wavelength range, col. (5): FWHM at the central 4′ in Flamingos-2 images and at the 30″ in T-ReCS images, col. (6): total exposure time for each filter image.

## 3. Results

### 3.1 Circumnuclear Structure

The H-band image of the central region of NGC 253 is shown in Figure 2, which depicts the bar structure, the bulge component, and the circumnuclear disk. The last one is examined in Y, J, H, and Ks bands in order to highlight the nuclear structure changes with wavelength (Figure 3). For reference, we have noted the astrometric positions of the TH2 radio source and the K-band continuum peak (or infrared core, hereafter IRC). The more important features observed in all bands are the three brightest knots in the center. Also, some plumes are detected emerging from the nuclear disk, probably associated with the known nuclear outflow (Bolatto et al. 2013). A group of diffuse knots is prominent in the northeast quadrant. Particularly in the Y-band image, the main knots appear much more diffuse than in the other NIR filters. Additionally, some dusty patches are clearly visible in the Y-band image. Some of these plumelike structures are still visible in J band but they are not clearly evident in the H and Ks bands. In the Y, J, and H bands, a tail-like structure is neatly seen to southwest of the IRC position. Notably, the bright knots of star formation are placed toward the NE side of the IRC, and one can observe a few weak knots in the SW side of the IRC.

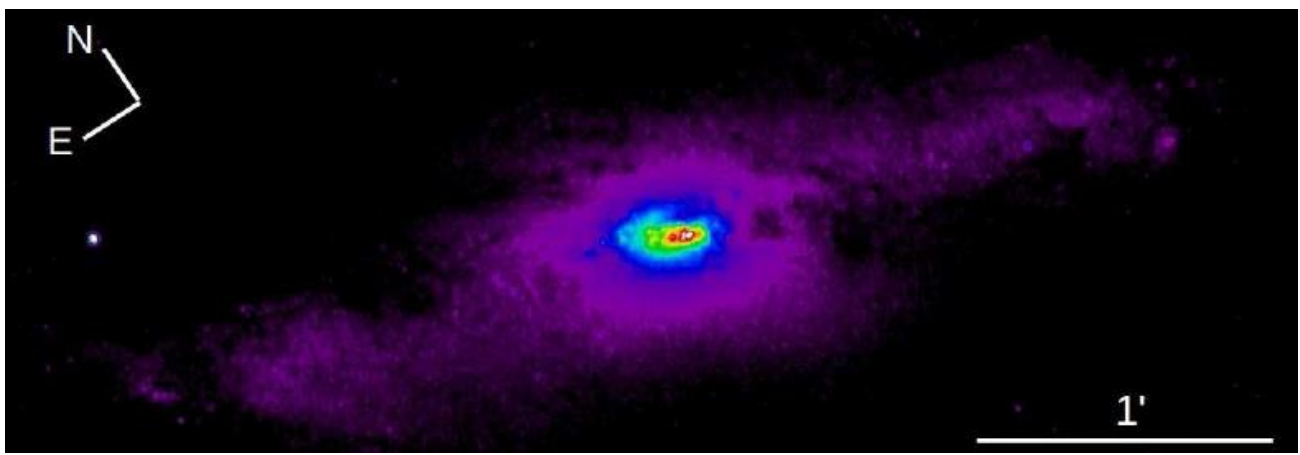

*Figure 2.* H-band image of NGC 253 displayed in logarithmic scale. This image includes the bulge and the bar of the galaxy. The bright knots in the very central zone are clearly seen.

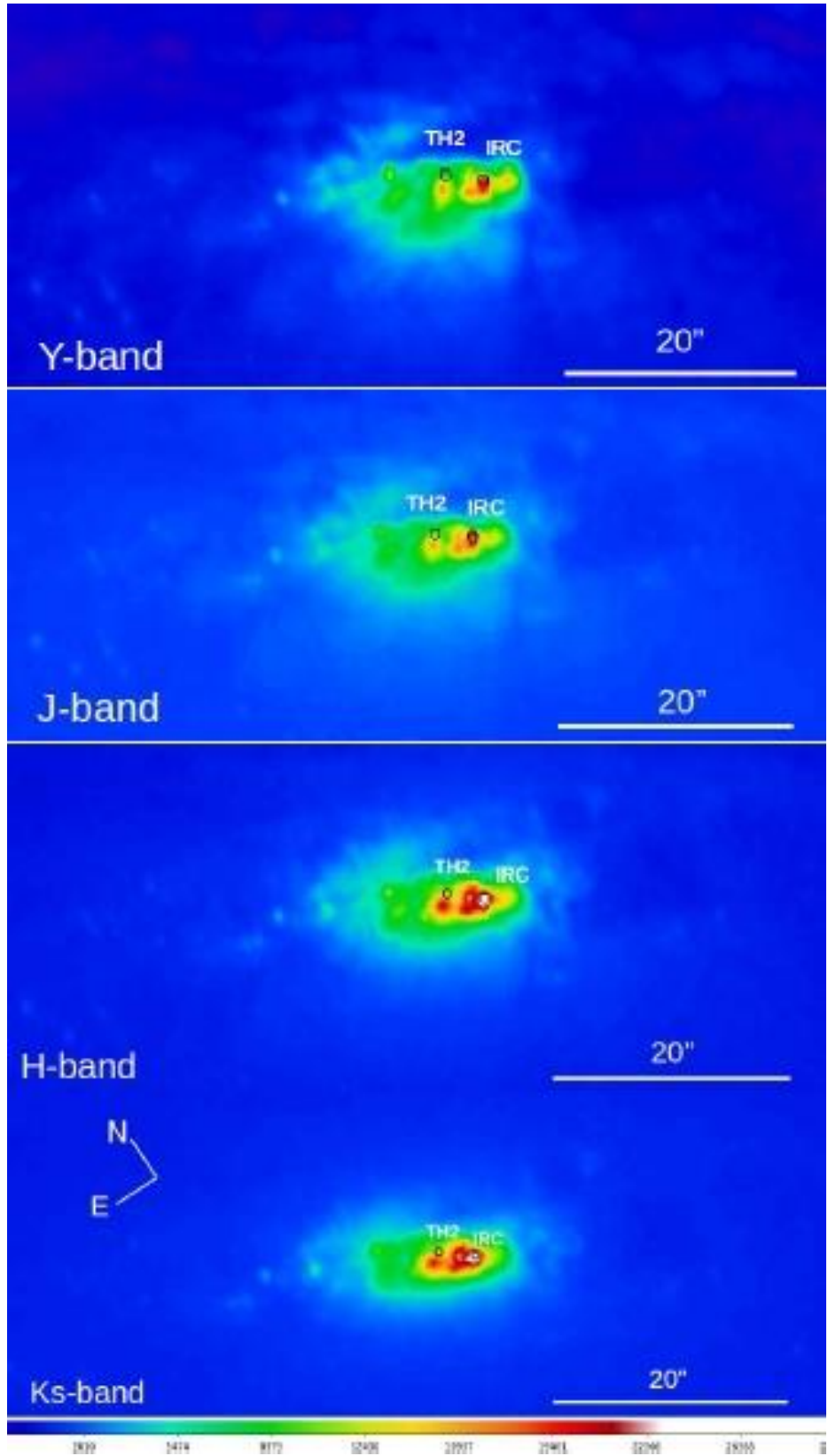

*Figure 3.* Y, J, H, and Ks band F2 images of the nuclear region of NGC 253. The positions of TH2 and of the K continuum emission peak (IRC) are noted.

The circumnuclear structure of NGC 253 in the MIR spectral range is presented in Figure 4. This pseudo-color TRecs image was built from the combination of the filters listed in Table 1 (the Si-2 filter is shown in blue, the [NeII] 12.8 μm in green, and the Qa broadband filter in red). The Si-2 filter at 8.7 μm reflects the presence of polycyclic aromatic hydrocarbon lines. [NeII] at 12.8 μm together with its adjacent continuum filter at 13.1 μm highlight emission line contribution, and the Qa filter at 18.3 μm traces flux originating from colder dust. However, Mentuch et al. (2009) have proposed that star-forming galaxies show an additional contribution of NIR and MIR flux originating in the hotter circumstellar dust around massive stars. The brightest MIR knot (IRC) is pointed out in the figure. The IRC matches with the Ks-band emission peak and the radio source TH7 (Fernández-Ontiveros et al. 2009). An evident structure observed in the composite image is a plume (in green) with a north-south orientation, which is centered on the TH2 position. In particular, the north side of the plume is clearly noted in [NeII] 12.8 μm and extends 25 pc down to a signal to noise of ~1. Southwest of the IRC, two faint green plumes are suggested. These structures are seen in [NeII] 12.8 μm and Si 8.7 μm. In the [NeII] continuum at 13.1 μm, they just appear faint and diffuse. No emission is seen in Qa band. An interesting arclike feature surrounding the IRC is observed in the continuum-subtracted [NeII] image (Figure 5). This feature cannot be due to hot dust, as it is not present in 13.1 μm and can confidently be linked to ionized gas emission.

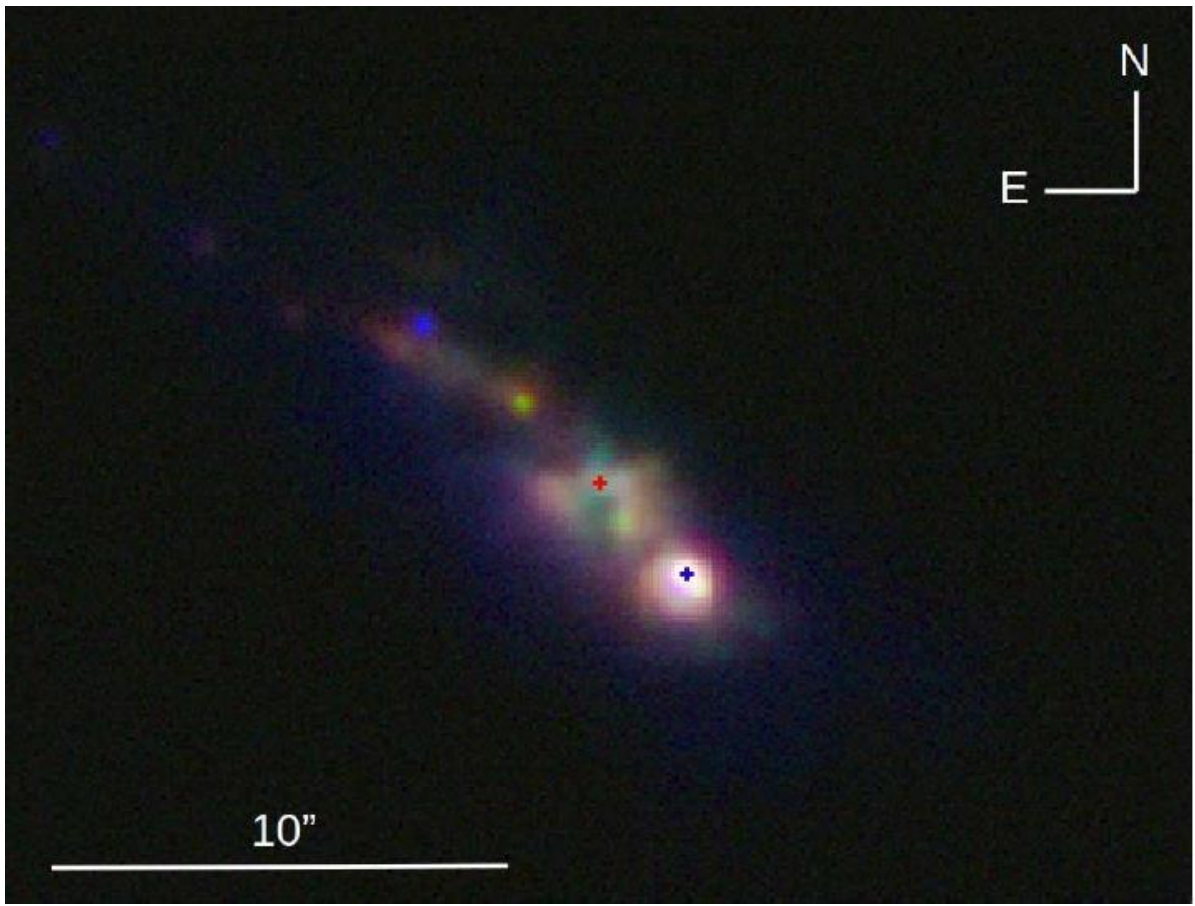

*Figure 4.* T-ReCS composite image of the nuclear region of NGC 253. The Si-2 filter is shown in blue, the [NeII] 12.8 μm in green, and the Qa broadband filter in red. The red cross indicates the position of TH2, while the blue one marks the position of the IRC.

The filamentary structures that seem to originate in the surrounding region of the IRC toward the SW are seen both in the T-ReCS filters ([NeII] and Si 8.7 μm) and in the Paα emission image. The HST NIR images of Alonso-Herrero et al. (2003) were analyzed in order to compare them with the T-ReCS MIR images. The main observed structures in the last ones were better recognized in the HST Paα filter image which was displayed in Figure 1. For a comparison, in Figure 5 (bottom) the

HST Paα filter contours were superimposed on the T-ReCS image composed by the addition of all filters listed in Table 1. It is necessary to note the resolution obtained with T-ReCS, which is of the same order of that achieved with HST.

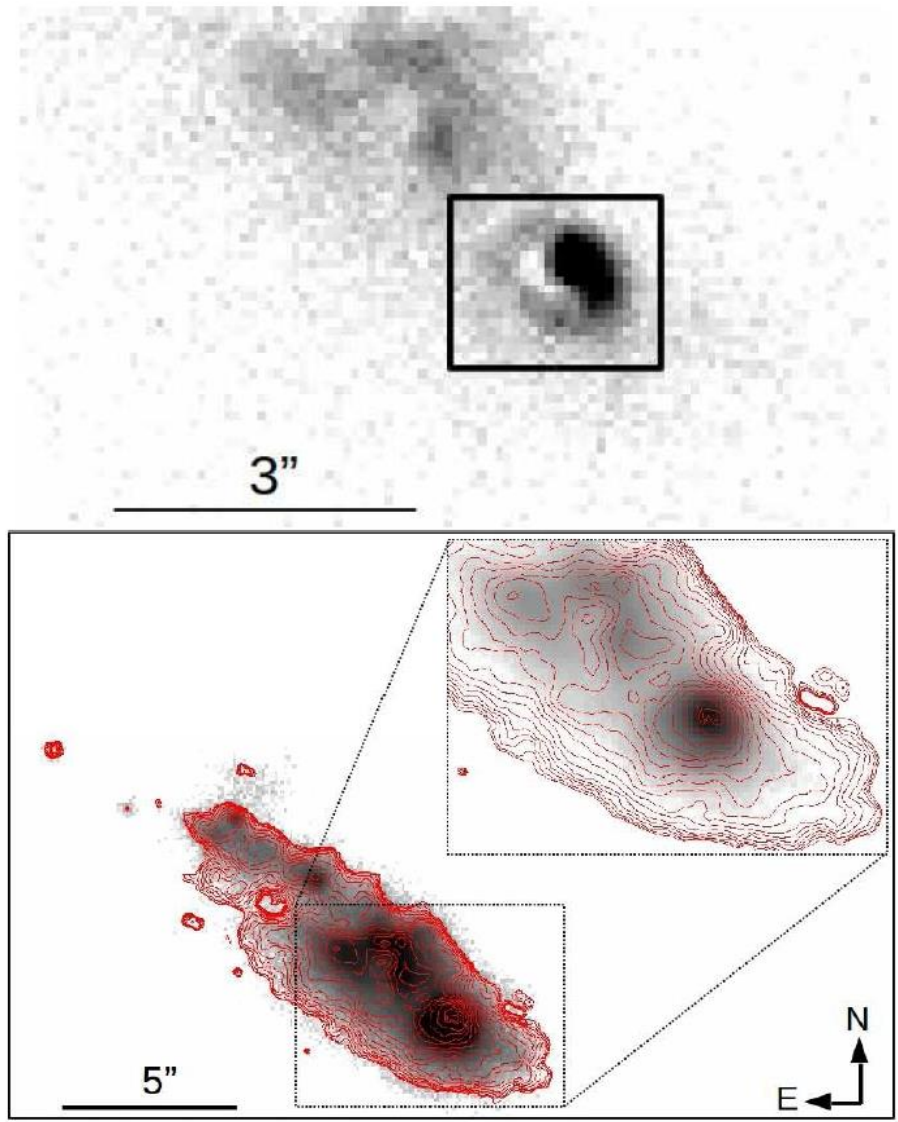

*Figure 5.* *Top: zoom of the continuum-subtracted [NeII] 12.8μm image. In the same image, the bubblelike structure is bounded by the rectangle. Bottom: image displayed in logarithmic scale, obtained from the addition of the images corresponding to the T-ReCS filters detailed in Table 1. The Paschen α isophotes are superimposed and, in the upper corner, a zoom is shown with a different contrast for highlighting the details around the IRC.*

In order to distinguish the arc structure and other MIR peculiarities, an unsharp masking Qa image was constructed (Figure 6). The most conspicuous regions or knot structures are pointed out and cataloged in Table 2. As in the NIR images, almost all the star formation knots are placed in the NE side of the IRC. The Qa fluxes were calibrated following the procedures established for T-ReCS data reduction cited on Gemini website. The Integration Time Calculator was used to estimate the uncertainties for differences that may exist in AB magnitudes considering the spectral energy distribution (SED) of sources (with the options varying between a starburst galaxy spectrum and a black body spectrum with a temperature of 200 K).

In the Figure 6 inset (3.6" x 2.6") the arc around the IRC is clearly visible. The IRC is cataloged as A1-1, which is a very compact source with FWHM ~0.4" (~6.8 pc) in size in the Si-2 8.7 μm T-

ReCS images. As can be seen, there is a substructure at ~7 pc SW from the IRC, which was cataloged as A1-2. It is at the end of the arc or shell. This arc spans around the brightest source, A1-1, and presents an inner radius of ~11 pc and an outer radius of ~16 pc. The best resolved image in the HST data set studied by Alonso-Herrero et al. (2003) is the one through the Pα filter. Kornei & McCrady (2009) identify an asymmetry in the region associated with IRC, which we resolve as a shell structure in the [NeII] T-ReCS image.

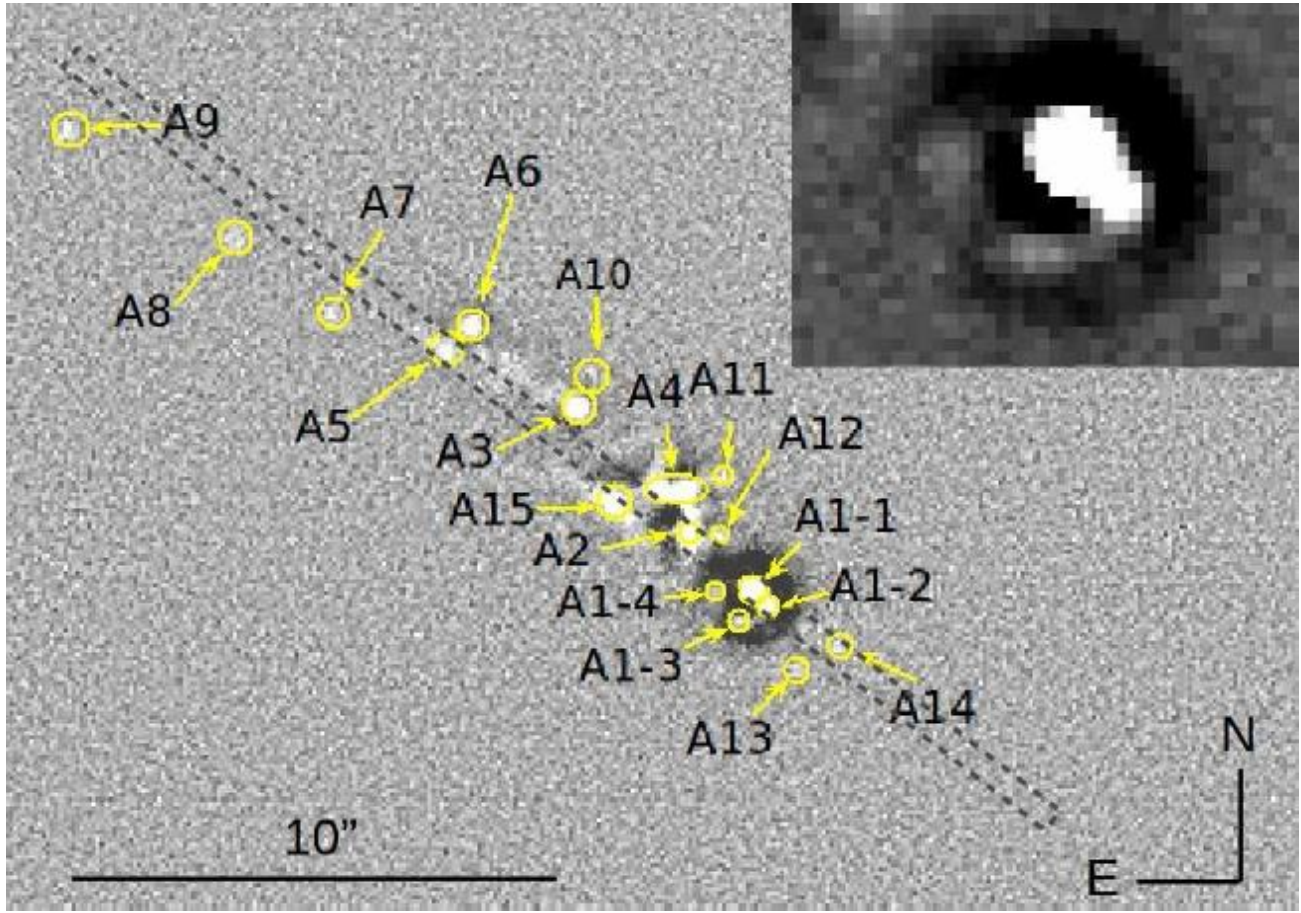

*Figure 6.* T-ReCS image obtained applying the unsharp mask technique. In the plot are indicated the different regions/knots identified. The position of the slit used for the F2 spectroscopy is also indicated. (Inset): Region that includes A1-1 (the region with the IRC peak as center), A1-2, and the arc structure. The inset is 3.6" x 2.6".

A knot that is very bright in Si 8.8 μm is A6 (IR32 in Fernández-Ontiveros et al. 2009), which is pretty much dimmer in the other MIR filters. On the other hand, region A3 is fainter in Si 8.8 μm band and prominent in the other filters, especially in the Qa band (see Figure 4). For reference, the position of the F2 slit is outlined in Figure 6.

In Figure 7, a Y-Ks color map is displayed where structures like the bar, the arms, and a nuclear starburst stand out and are seen with high resolution. The most notable features are the known dust filaments which depart from the starburst region in radial directions. These filaments were reported by Sugai et al. (2003) observed in HST images mapping the $H_2$ 1–0S(1) emission. They distinguish four filaments which they called A, B, C, and D, of which both of those that were located in the Southeast direction (A and B) are recognized in the F2 color map (Figure 7). Sugai et al. (2003) report a filament extension of 10″ (170 pc) while these features extend up to ~ 260-370 pc in the F2 images. Additionally, a third filament toward the west of filament A can be distinguished in the F2 color map. However, the filaments C and D (in the northwest) were not observed in any F2 filter or color map image. The circumnuclear region is displayed in Y-Ks, J-Ks, J-H and H-Ks color maps (Figure 8). It can be seen that the IRC belongs to a region of high reddening. The most distinctive structure in these color maps is a disk-like feature probably associated with a starburst. The origin of the longest filament (visible in every color) is near the symmetry center of this mentioned structure. Both the IRC and TH2 are placed at the northeastern side of the color disk.

Table 2
NGC 253 MIR Emission Regions

| Region | $\alpha$ (J2000) | $\delta$ (J2000) | Position Error (Radius) ('') | Qa Fluxes (mJy) |
|---|---|---|---|---|
| A1-1 | 00:47:32.94 | −25:17:19.70 | 0.08 | 8272(A1-1+A1-2) |
| A1-2 | 00:47:32.93 | −25:17:20.02 | 0.08 | 8272(A1-1+A1-2) |
| A1-3 | 00:47:32.96 | −25:17:20.37 | 0.09 | <10 |
| A1-4 | 00:47:33.00 | −25:17:19.75 | 0.10 | <10 |
| A2 | 00:47:33.05 | −25:17:18.42 | 0.09 | 1303 |
| A3 | 00:47:33.22 | −25:17:15.74 | 0.08 | 45 |
| A4 | 00:47:33.05 | −25:17:17.52 | 0.10 | 1215 |
| A5 | 00:47:33.43 | −25:17:14.50 | 0.11 | 248 |
| A6 | 00:47:33.38 | −25:17:13.96 | 0.07 | <10 |
| A7 | 00:47:33.62 | −25:17:13.70 | 0.11 | 50 |
| A8 | 00:47:33.76 | −25:17:12.18 | 0.11 | 55 |
| A9 | 00:47:34.03 | −25:17:09.69 | 0.09 | 28 |
| A10 | 00:47:33.19 | −25:17:14.94 | 0.12 | <10 |
| A11 | 00:47:33.00 | −25:17:17.26 | 0.12 | <10 |
| A12 | 00:47:32.99 | −25:17:18.49 | 0.11 | <10 |
| A13 | 00:47:32.87 | −25:17:21.53 | 0.11 | <10 |
| A14 | 00:47:32.81 | −25:17:20.91 | 0.11 | <10 |
| A15 | 00:47:33.17 | −25:17:17.73 | 0.10 | 1305 |

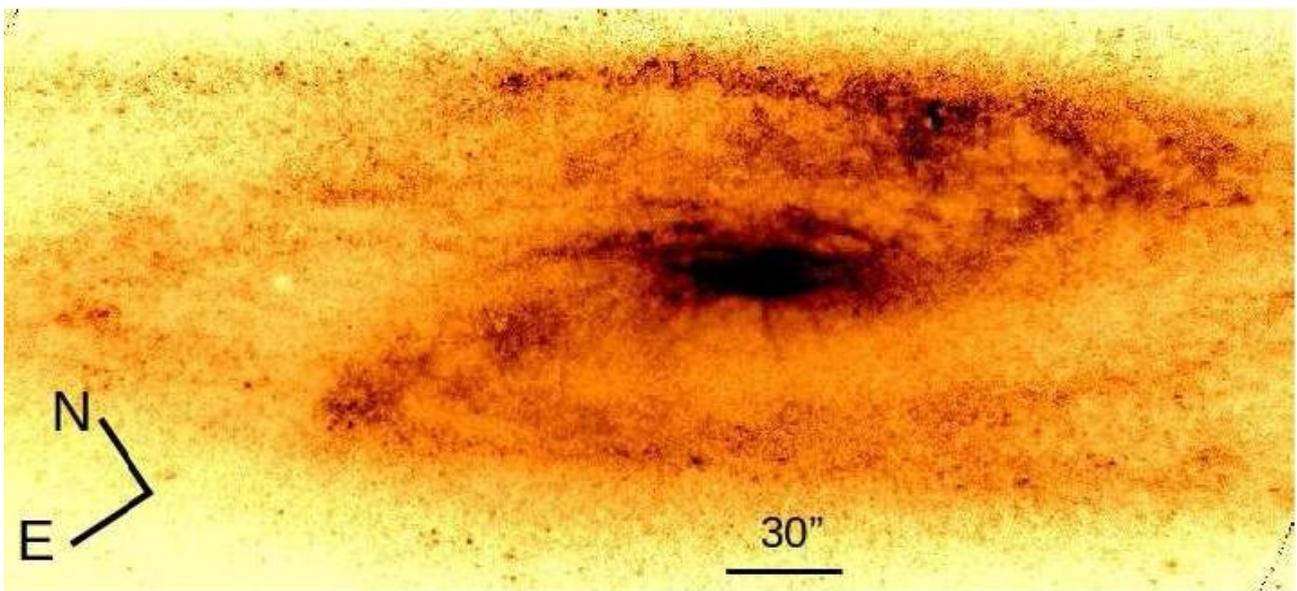

*Figure 7.* Y-Ks color map. The star formation regions are highlighted and the prominent features observed in the map are the known outflow filaments emerging from the circumnuclear disk.

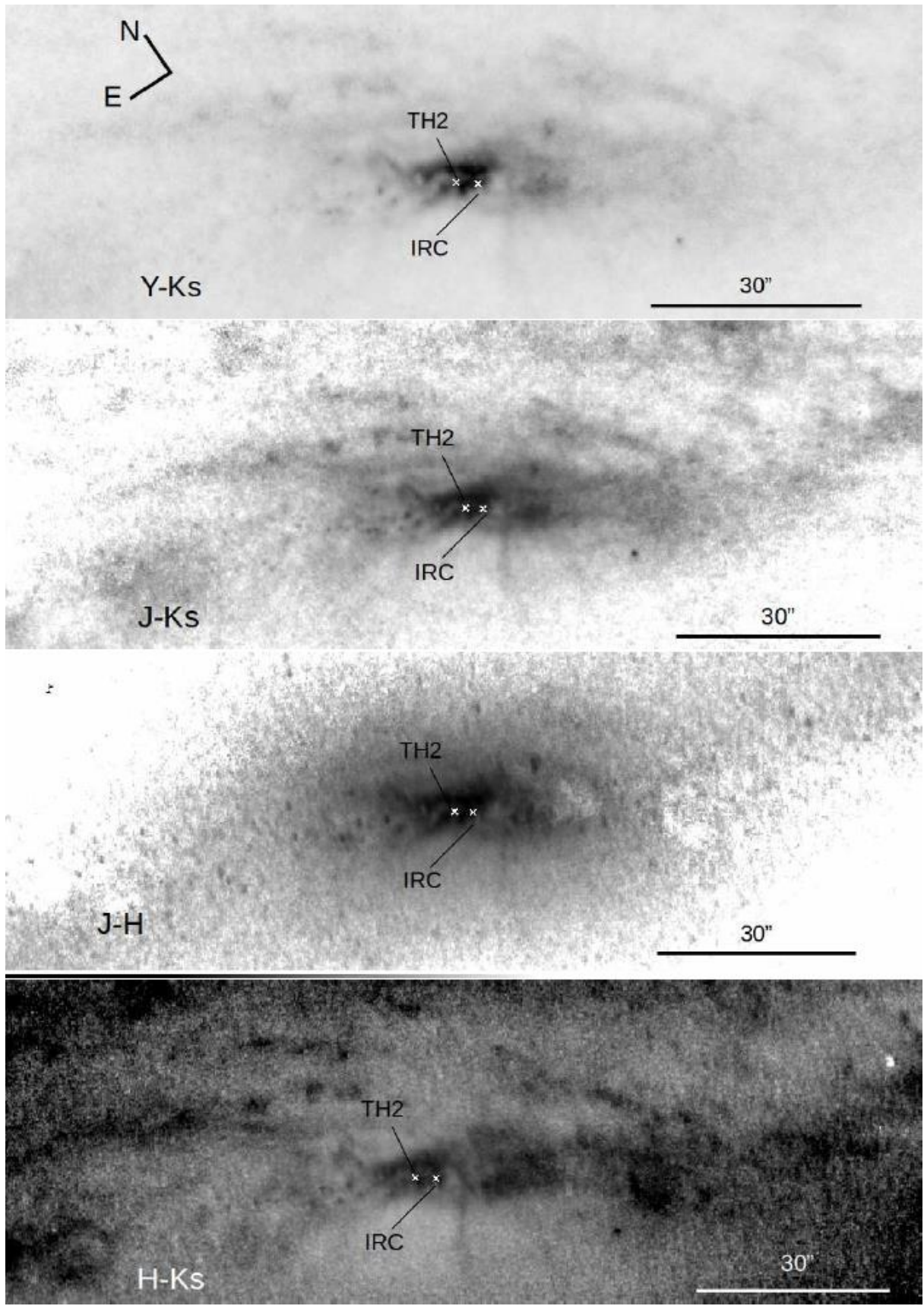

*Figure 8. Y-Ks, J-Ks, J-H and H-Ks color maps. Note the positions of TH2 and the NIR continuum emission peak (IRC); the latter is the object nearest to the center of symmetry of the circumnuclear disk. The darkest zones correspond to a deeper reddening.*

## 3.2 Inner Structure

We start with a general morphological description of the central region using the broadband images for characterizing the largest angular feature, which is the bar, and in the next section we use the spectroscopic continuum profile to measure the innermost morphological substructure and compare it with its kinematic properties. A detailed photometric study of the galaxy bar and inner ring regions will be presented in a future paper (Camperi et al. 2016, in preparation).

We use the Ks band data to trace the mass structure. These data are the least affected by dust and they are dominated by the emission of the old stellar population. Thus, we estimate the bar position angle (P.A) in the Ks image, P.A. 71.6°, and derive the Ks band radial profile along the bar direction (Figure 9, above). In order to have an adequate sampling of the underlying structures, a slice of 16.3" wide was used. As a consequence, the shape of the radial profile is an average of the brightness along the slice. In the profile, we identify the bulge component with the highest Ks brightness (~11.4 mag arcsec$^{-2}$), the bar component with nearly constant Ks magnitude (~15.4 mag arcsec$^{-2}$ in the Ks image in accordance with Iodice et al. 2014), and we identify the disk with an outward linear decay in the same plot. The origin of the radial positions is selected in coincidence with the IRC position. We are interested in determining the center of symmetry of such structures. In order to identify them, the bar and the disk were characterized by straight lines in the logarithmic plot. In addition, the bulge component was fitted with a surface brightness $r^{1/4}$ profile (de Vaucoulers 1959). These component representations are plotted in Figure 9 (above). We consider that the bar ends where the brightness decays in half a magnitude. The bar length resulted in ~302" (deprojected diameter) with a center of symmetry located at 1.4" (2.0" deprojected position) toward the NE from the IRC. The disk is well described with a scale length of 28.6" = 486 pc with a center of symmetry placed 1.4" towards East from the IRC, consistent with the bar center. However, the symmetry center of bulge component is positioned on 2.6" (3.7" deprojected position) east of the IRC.

We can also characterize the morphology by using the radial color profiles. In the Y-band image, the star formation regions and dust lanes stand out in contrast to the Ks band image, which is a lot more uniform and it is considered less sensitive to star formation regions. Consequently, Y-Ks color is the most suitable for tracing structures like the bar and spiral arms (Figure 9, bottom). In that profile, the bar is identified with the region that oscillates around an average magnitude color of around 1.17, while the disk has a pronounced outward decreasing color. The central color structure is associated with the disklike feature revealed in the circumnuclear color map (Figure 7), which is probably related to the circumnuclear starburst process. We assume that this color structure has a linear increasing color toward the center with a symmetry center at 4.8" (6.9" deprojected position) towards SW of the IRC. In the same plot, the peaks of color at the ends of the bar are pointed out, associated with the ansae features. The bar ansae centroid positions have a center of symmetry at 1.5" NE from IRC, almost coincident with that obtained from the Ks-band bar profile (1.4" NE of the IRC).

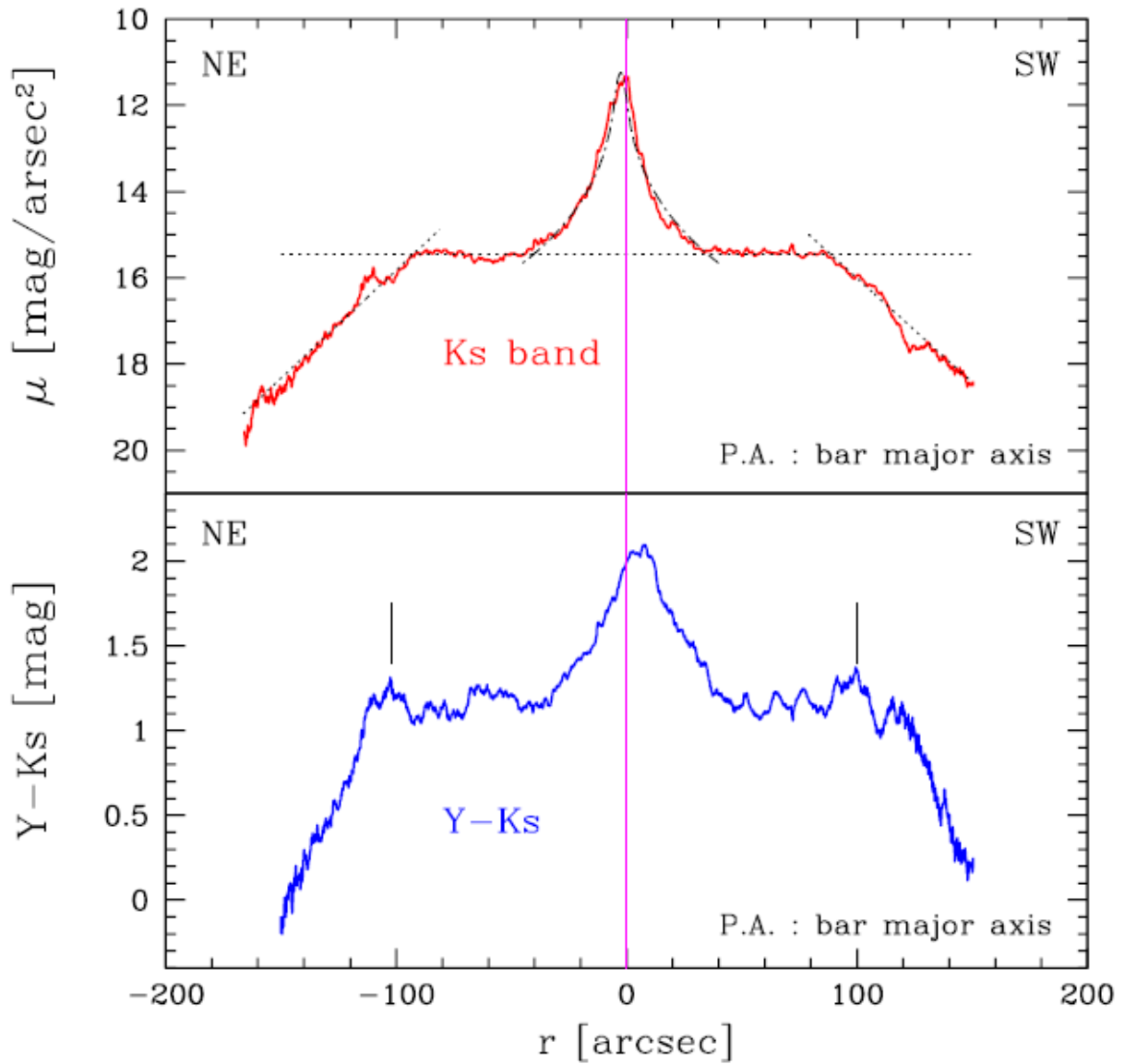

*Figure 9.* Top: Ks band magnitude profile along the bar major axis. In the upper plot we normalized the values that correspond to the peak surface brightness, that is, those values that would result from about a 1 pixel width extraction. The origin is set at the IRC position. The bulge component is fitted with a dotted-dashed line and the disk fitting with dotted lines. Bottom: Y-Ks color profile along the bar major axis.

### 3.3. Spectroscopic Radial Profiles

Figure 10 shows an NIR spectrum of the nuclear region (IRC) and of a region 7.1" NE from the IRC. The most conspicuous emission lines are identified, and the change in the relative intensity between Brγ and $H_2$ emission lines is evident.

The Brγ line profile (Figure 11) has a marked blue wing (blueward asymmetry) in the 10" IRC surroundings. These line asymmetries extend toward the NE (5.4") and the SW (4.7") of the IR peak and are less prominent with distance from the IRC. This could be associated with an outflow process. However, the $H_2$ molecular lines have symmetric profiles in the whole extension (Figure 11). On the other hand, the HeI emission line, although much noisier, presents the same asymmetry as the Brγ emission line. This spectral feature extends along the whole spatial range. For comparison, a sky emission line profile is included in Figure 11 which results in it being highly symmetric (FWHM ~ 0.5").

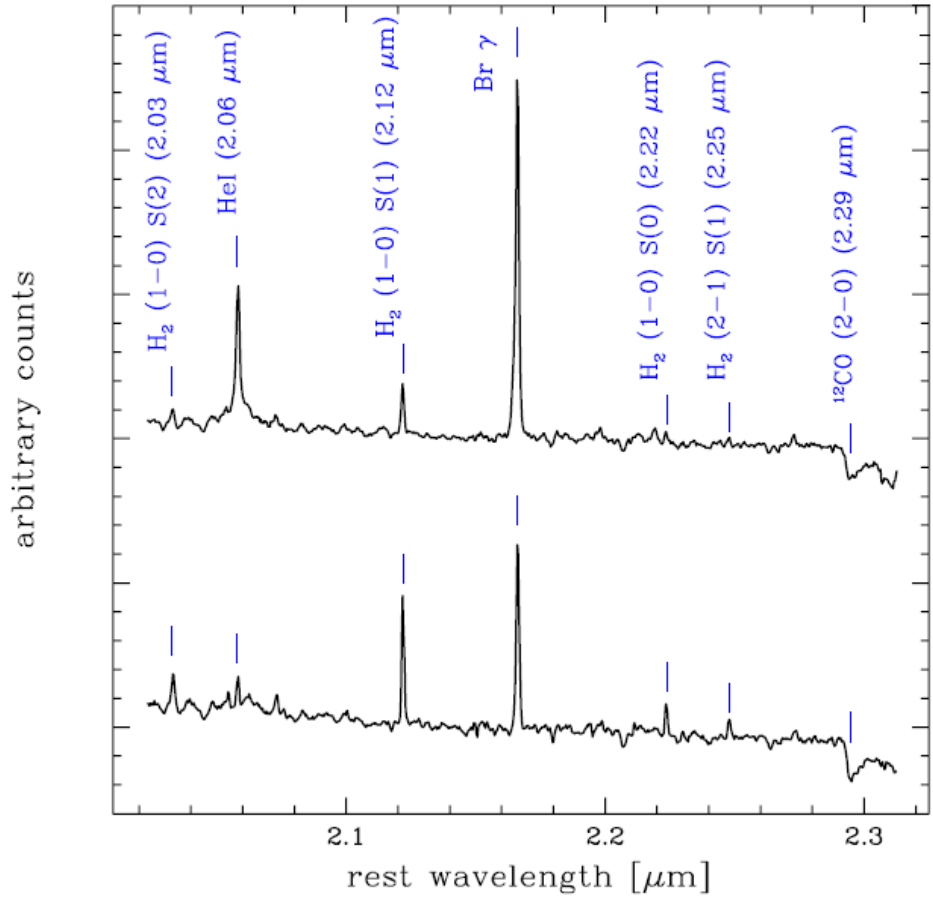

*Figure 10.* Top: NIR spectrum corresponding to region A1-1. Bottom: Off-nucleus spectrum, about 7.1"
northeast of the IRC.

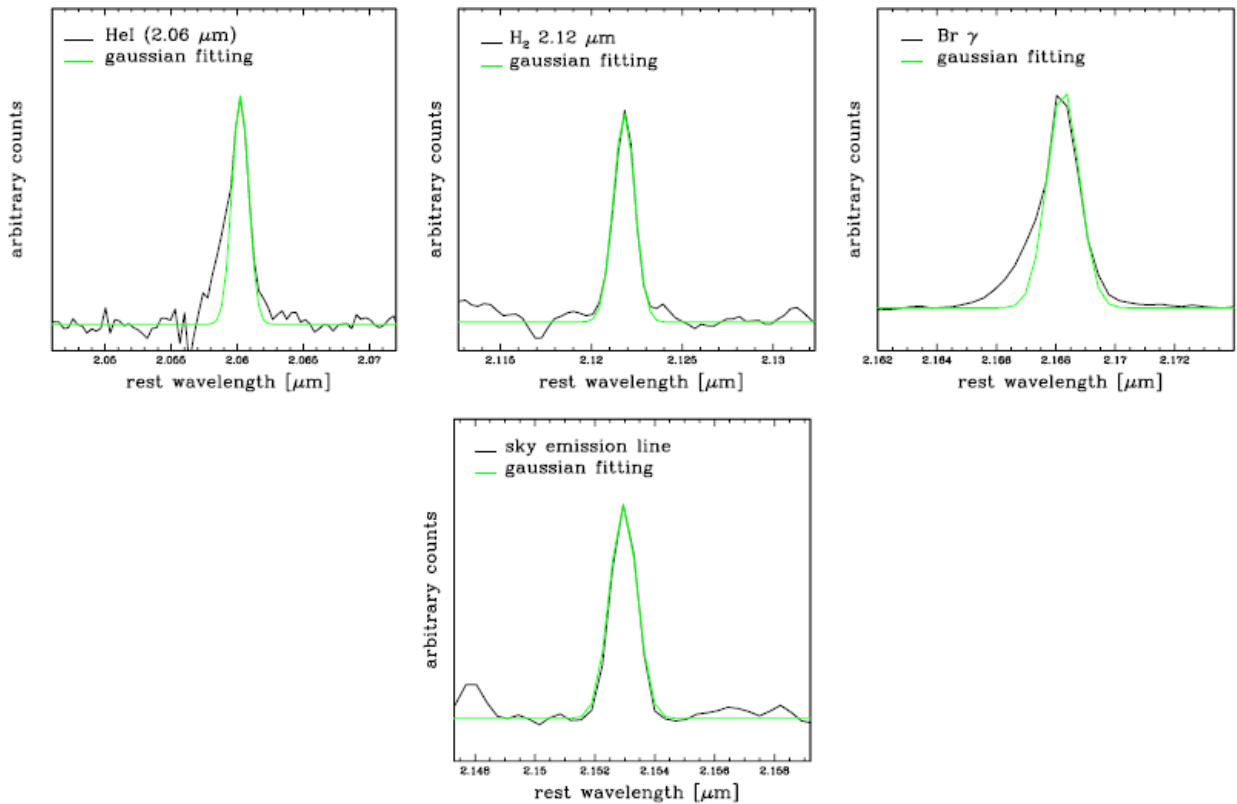

*Figure 11.* Top left : HeI 2.06 μm asymmetric emission line profile. Top center: H$_2$ 2.12 μm symmetric
emission line profile. Top right: Brγ asymmetric emission line profile. Bottom: sky emission line profile.

We have obtained the continuum emission spatial profile next to the 2.12 μm H$_2$ emission line with a spectral width of 0.02 μm. The continuum emission peak is coincident with the position of the brightest infrared source (IRC) with an uncertainty of 0.18". Additionally, we have analyzed the radial profiles of the main emission lines present in the spectrum arising in the neutral hydrogen (Brγ), neutral helium (HeI 2.06 μm), and molecular hydrogen (H$_2$ 2.12 μm). In Figure 12, the spatial profiles for H$_2$, Brγ, He I, and 2.12 μm nearby continuum along the slit are plotted.

The main MIR knots of Table 2, along this position angle (PA) have also been identified. These spatial profile plots show that the main infrared source, the IRC, is the dominant nuclear source in the NIR. Hereafter, we take this IRC peak as the origin for radial positions. NE of the IRC we detect several secondary intensity peaks which decay in intensity away from the IRC. Toward the SW the intensity decays abruptly, as was noted in the circumnuclear NIR images.

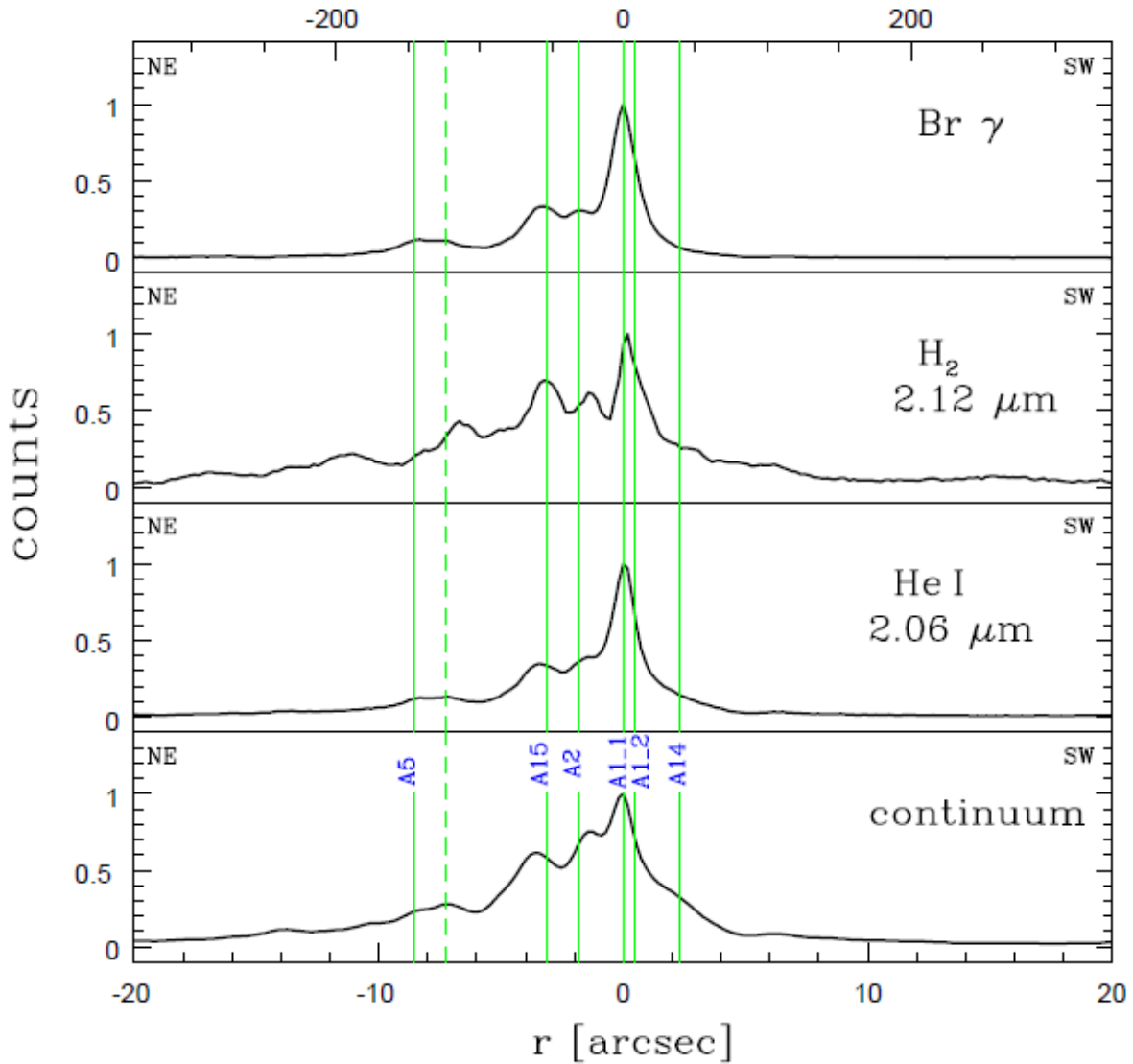

*Figure 12. Emission lines and continuum radial profiles. For comparison, the positions of the knots cataloged in the T-ReCS infrared images are plotted with green lines. The dashed line corresponds to a position coincident with a nebulosity-like region placed nearly between A3 and A5. (This zone does not match any of our cataloged regions).*

The continuum profile is strongly dominated by the old stellar population, in contrast to the Ks broadband profile which includes other emission sources. For that reason, the continuum profile allows us an exhaustive analysis of the innermost circumnuclear structures. Figure 13 (above) shows the continuum emission profile over the central 2′. In the outermost region, the underlying structure was fitted with a surface brightness $r^{1/4}$ profile (de Vaucoulers 1959). The fitting was tuned to minimize the residuals at a radial range of 20" - 70". The symmetry center of this luminosity component resulted in a location 7.4" to the NE of the IRC position. The scale length of the fitted bulge is (10.5±0.9) pc, using a Gaussian convolution of 0.5" in order to account for the seeing effect. This stellar bulge was subtracted from the continuum profile, leaving a residual continuum that exhibits two other stellar structures (Figure 13, bottom). One of them is more extended and less luminous. Its center coincides with the secondary peak placed at 3.62" NE of the IRC and it was represented by a Sérsic law with an index of 0.5 (for a description of Sérsic's profile, see Binney & Merrifield (1998)). The other component is sharp, compact and more luminous. It was represented by an exponential profile (disk profile) with a center at 0.5" NE of the IRC and a scale length of 30 pc. Both components and the total residual fitting are shown in Figure 13 (bottom). This stellar structure description will be useful in modeling the 2D galaxy luminosity in a future paper (Camperi et al. 2016, in preparation).

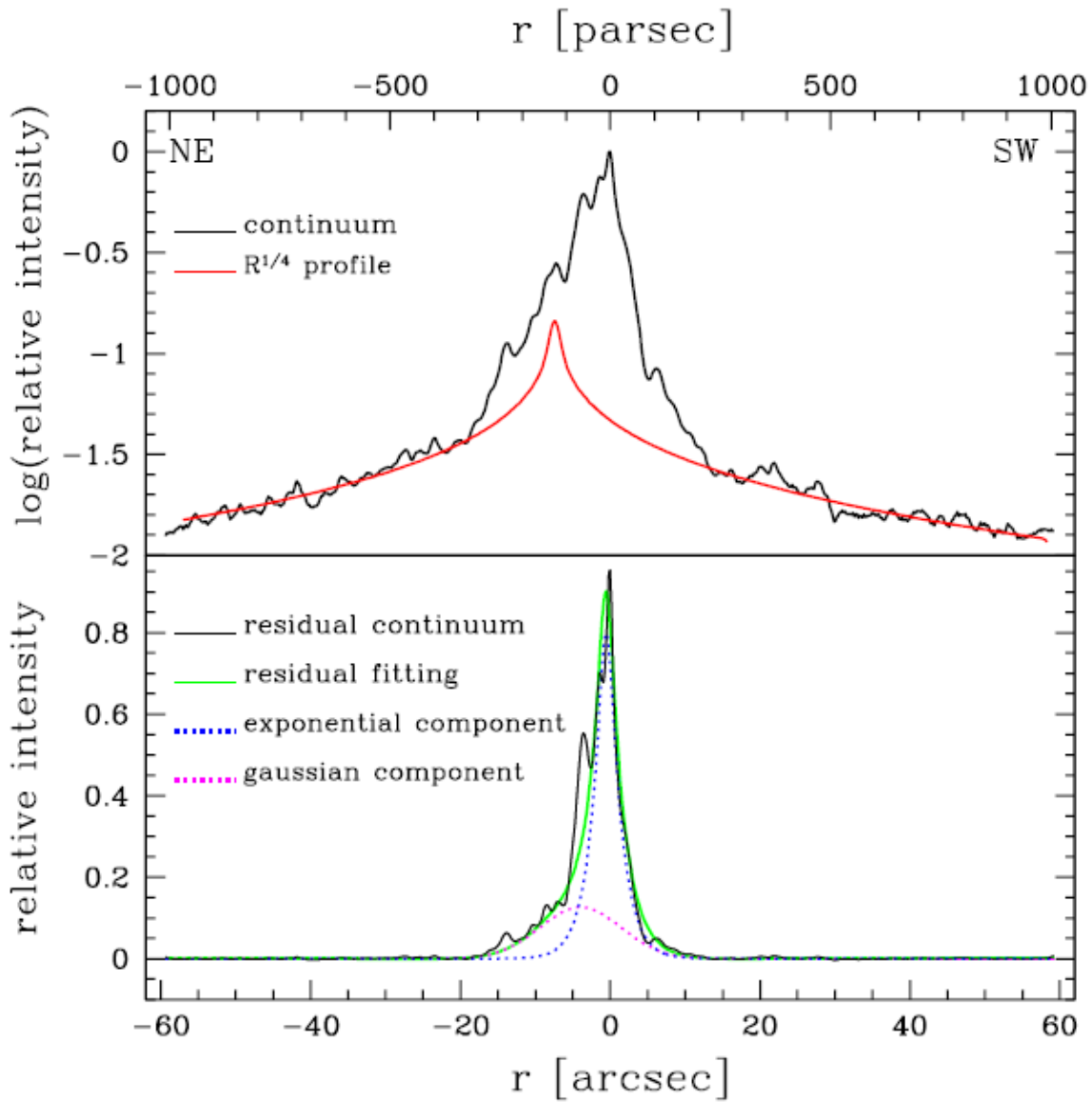

*Figure 13.* Top: continuum radial profile next to the Brγ and $r^{1/4}$ law fitting of a bulge component. Bottom: residual continuum profile ($r^{1/4}$ model subtracted) and fitting of this residual (in green). The residual is composed of two components: an exponential disk (blue dots) and a Sersic (n=0.5) component (magenta dots).

## 3.4. NIR Diagnostic Diagram

NIR emission lines can be used as diagnostics of the physical state of the $H_2$ gas (Falcón-Barroso et al. 2014). To achieve this goal, we followed a number of previous studies (Mouri 1994; Rodriguez-Ardila et al. 2004, 2005; Falcón-Barroso et al. 2014). These studies exploit the fact that in the interstellar environment, most of the lowest vibrational levels ($v = 1$) of $H_2$ tend to be well thermalized, while higher level transitions are principally populated by processes such as non-thermal UV fluorescence. The H2 molecule, generally speaking, can be excited by two kinds of mechanisms with subsequent thermal and non-thermal emission (Reunanen et al. 2002; Falcón-Barroso et al. 2014):

(1) Thermal mechanisms can be classified as (a) shocks, gas motions at very large velocities that accelerate and heat the ambient gas (e.g. Hollenbach & McKee 1989), (b) UV radiation in dense (n $\geq 10^4$ cm$^{-3}$) clouds (Sternberg & Dalgarno 1998), and (c) X-ray illumination, which is the name given to the processes where hard X-ray photons penetrate deep and heat large amounts of molecular gas (e.g. Maloney et al. 1996).

(2) The non-thermal mechanism involves UV fluorescence in interstellar media with low density. In this case, photons with $\lambda > 912$ Å (the ionization threshold) are absorbed by the H2 molecule and then re-emitted (e.g. Black & van Dishoeck 1987). The 1–0S(1)/2–1S(1) line ratio has showed itself as a great discriminator between thermal and non-thermal processes. Following from the models of Mouri (1994), this ratio takes lower values ($\leq 2$) in the regions that are dominated by UV fluorescence and rises to larger values in the thermal-dominated regions ($\geq 5$). Given that both lines have similar wavelengths and are independent of the ortho/para ratio, a diagnostic of this kind has the advantages of being fairly insensitive to extinction. However, the 1–0S(2)/1–0S(0) line ratio is sensitive to the strength of the incident radiation and so it can be used to discriminate the dominant excitation process.

Following Falcón-Barroso et al. (2014), we show in Figure 14 the value of the flux ratio of the 1–0S(1) and 2–1S(1) versus the ratio of the 1–0S(2) and 1–0S(0) lines. We performed spectrum extractions of selected zones according to the most conspicuous features in the continuum profile. The positions and widths of these regions are detailed in Table 3. The IRC belongs to Region 1 and TH2 to Region 3. In Figure 14 a dotted-dashed line is shown, which corresponds to the vibrational temperature ($T_{vib}$) equal to the rotational temperature ($T_{rot}$) of the $H_2$ gas, which was calculated using the expressions in Reunanen et al. (2002). In the case of thermal excitation, both temperatures should be similar (i.e. close to the dashed line in Figure 14), whereas in the case of non-thermal excitation, $T_{vib} \gg T_{rot}$. For all these regions, the main excitation mechanism is thermal. As can be seen, region 1 (which surrounds the IRC) is the most thermalized one and we could conjecture that this region could exhibit a high density plus the action of strong shock waves.

Table 3
Spectrum Extraction Data

| Region | Position (″) | Extraction Width (″) | MIR Associated Regions |
|---|---|---|---|
| 5 | −8.3 | 1.1 | coincident with A5 |
| 4 | −7.1 | 1.5 | diffuse or nebular zone between A3 and A5 |
| 3 | −3.5 | 3.2 | includes A15 and, marginally, A2 |
| 2 | −1.2 | 1.5 | includes adjacent region of A1-1 and, marginally, A2 |
| 1 | 0 | 0.9 | includes A1-1 |
| 6 | 6.2 | 1.6 | without associated region in MIR |
| 7 | 20.4 | 1.8 | without associated region in MIR |

**Note.** Col. (1): region number in the NIR diagnostic diagram (Figure 14). All the extractions correspond to the P.A. = 51°. Col. (2): position in arcsec, taken as the origin of the IRC; negative values correspond to extractions toward the NE side. Col. (3): width of the spectrum extraction, in arcsec. Col. (4): MIR T-ReCS counterparts.

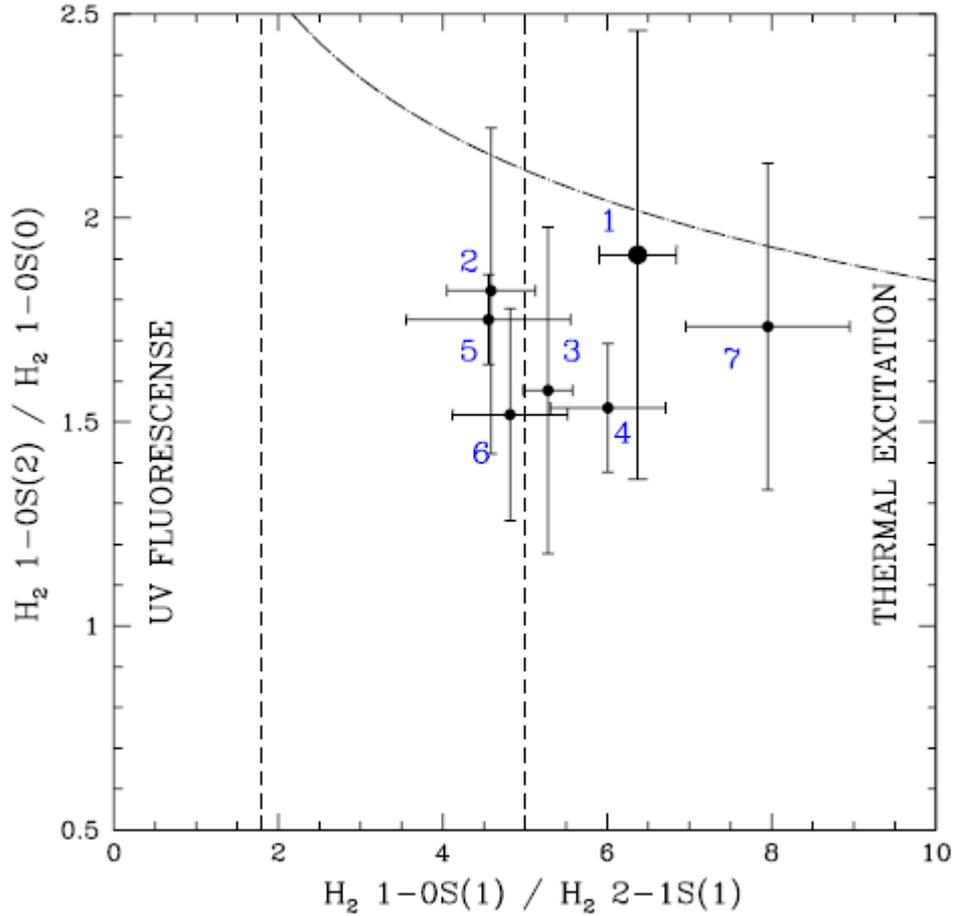

*Figure 14.* NIR diagnostic diagram. The numbers indicate the regions listed in Table 3. Note that region 1 is centered on the IRC and includes the MIR source A1-1.

**3.5 Brγ Equivalent Width (EW) vs. Brγ FWHM**

In Figure 15, we plot the star-forming indicator (EW(Brγ)) as a function of the width of the emission line (FWHM(Brγ)). The FWHM has been corrected by the instrumental width measured at the nearest atmospheric OH emission lines (9.5 Å). As can be appreciated, in general, the lowest values (lower than 10 Å for the EW) for both indicators have been found in the outermost regions (green dots; 11.9" to 17.7" in the NE direction and 5.8" to 9.0" in the SW direction). For intermediate regions, in the NE direction (blue dots; 2.9" to 11.6") the FWHM(Brγ) values show grater dispersion range (very low values from ~2 to ~13.5 Å) and the EW(Brγ) is generally higher than that for the outer regions. The SW intermediate region (cyan dots; 1.4" to 5.4") presents EW(Brγ) lower than ~15 Å and has a lower dispersion in FWHM than the central region. The innermost region (magenta dots) displays the highest values in both EW and FWHM. The IRC region presents the highest values in star formation index (EW(Brγ)~40 Å) and at the same time the most dynamically active region. Two blue points with the highest FWHM values correspond to positions 2.9" and 3.2" from the IRC near the position of TH2 (3.3" from the IRC). The point with the highest FWHM (18 Å) corresponds to a position 0.7" NE of the IRC. This position would be inside the shell/arc mentioned in Section 3.1.

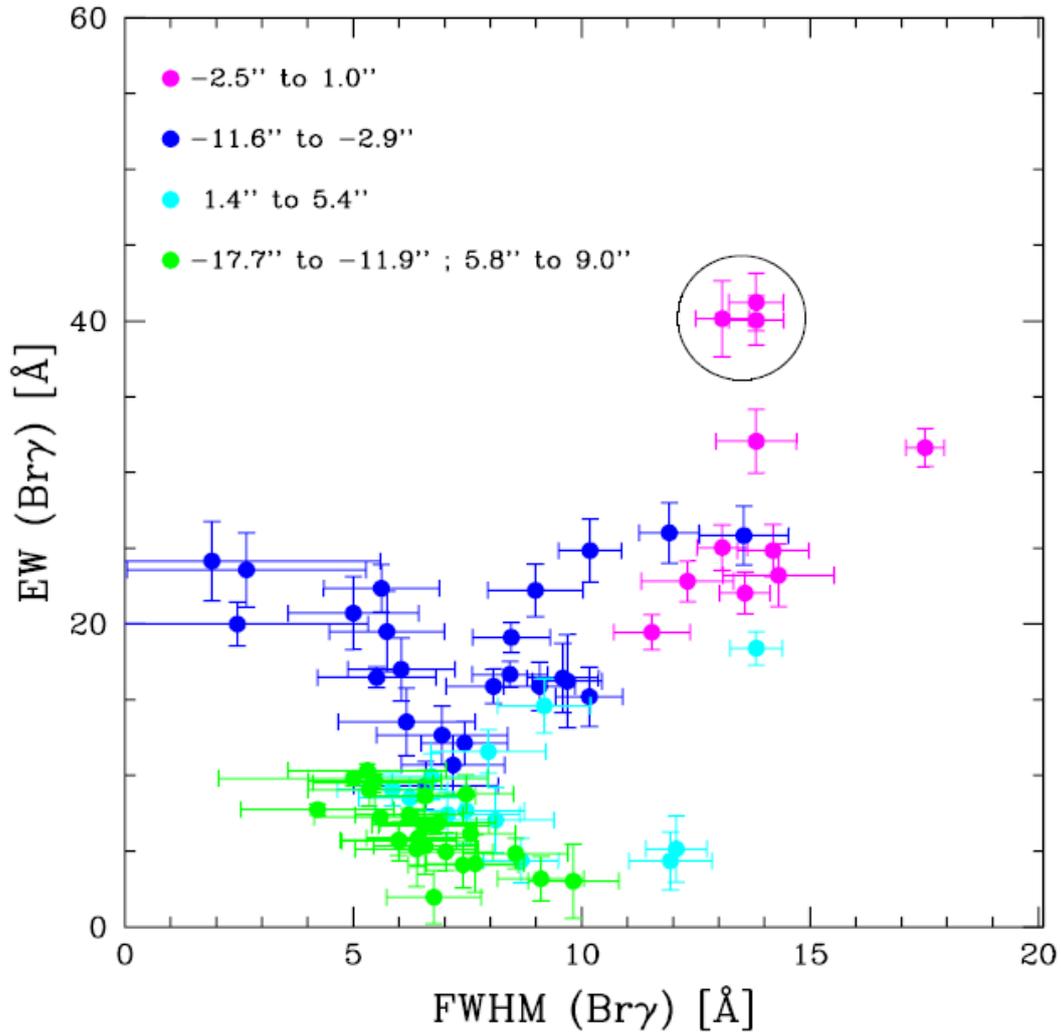

*Figure 15.* FWHM vs. EW of the emission line Brγ, corresponding to the regions across the slit. The different colors for the squares in the plot are assigned to three distinct regions: external (green), intermediate (blue), and central (red) regions. The radial distance range is specified in the plot, with the origin in the peak of the continuum emission (IRC). The three points inside the circle belong to the A1 region, the innermost region.

### 3.6 Kinematics

We have obtained the heliocentric radial velocity curves from Brγ and HeI 2.06 μm recombination emission lines and from the molecular hydrogen emission line ($H_2$ 2.12 μm). Both recombination lines showed asymmetry profiles in many locations (Figure 11). Thus, we have chosen the $H_2$ radial velocities to sample the rotation curve and therefore the mass distribution. This choice offers several benefits. First, the $H_2$ radial velocity measurements, when sampled in different individual spectra, have lower uncertainties and, as a consequence, the $H_2$ radial velocity curve is further well-behaved. Second, the cold molecular gas is more related to the thin circumnuclear disk and it appears less affected by radiation pressure and gas flows. Finally, the $H_2$ radial velocity curve presents the largest spatial range (about 50" or 850 pc).

The individual spectra were extracted every 0.36" (2 pixels) with an aperture of 0.36". Additionally, the central 5" of the radial velocity curve have been sampled with a slice thickness of 0.18", every 0.18" (1 pixel). The $H_2$ radial velocity distribution is shown in Figure 16 where the origin in radial position was adopted in coincidence with the position of the IRC brightness peak. The central 5" of the radial velocity curve show an S-shaped structure that is displayed in a zoom region in Figure 16. The central S-shaped perturbation is also present in the HeI and Brγ rotation curves. Most of the central rotation behaves like a solid body of about 22" in length (crosses in Figure 17).

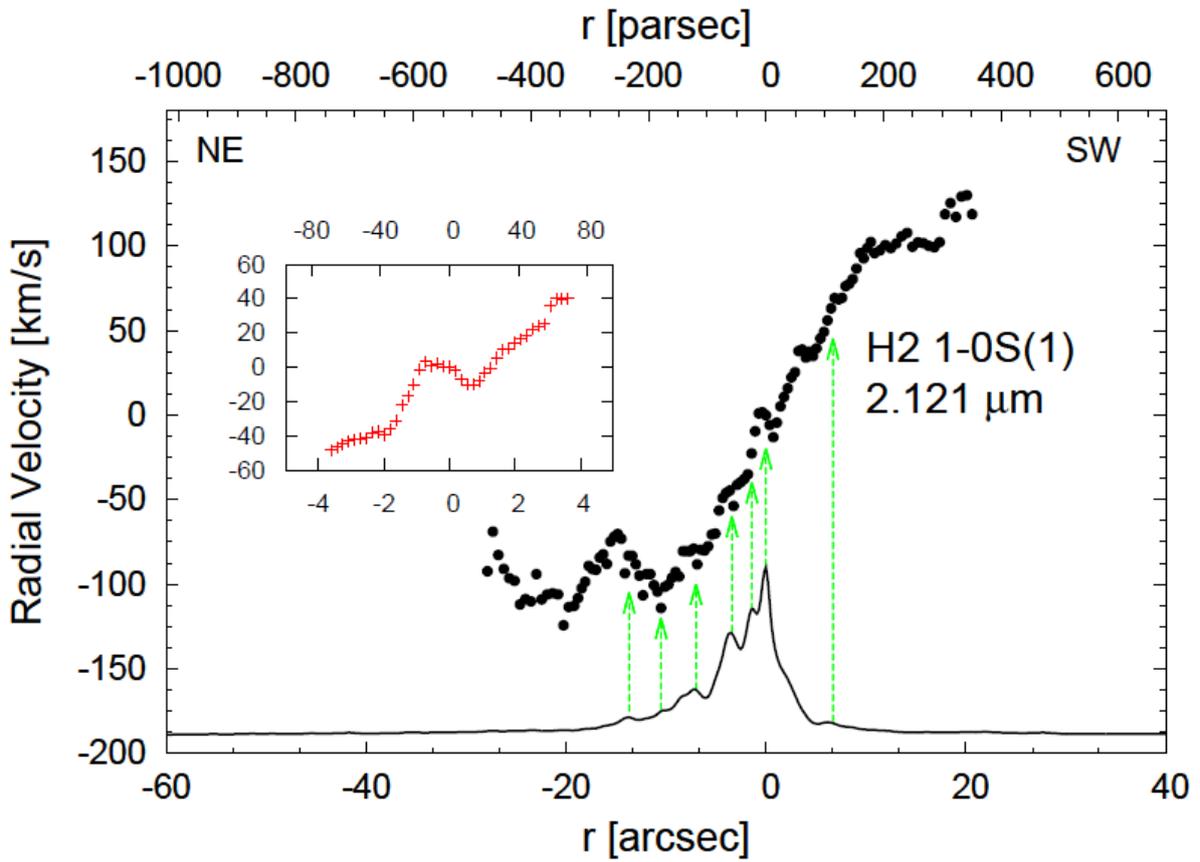

***Figure 16.*** $H_2$ *radial velocity distribution along the major axis. At the bottom of the figure the continuum profile next to the $H_2$ line is plotted. (Inset) zoom of the radial velocity distribution of the nuclear region.*

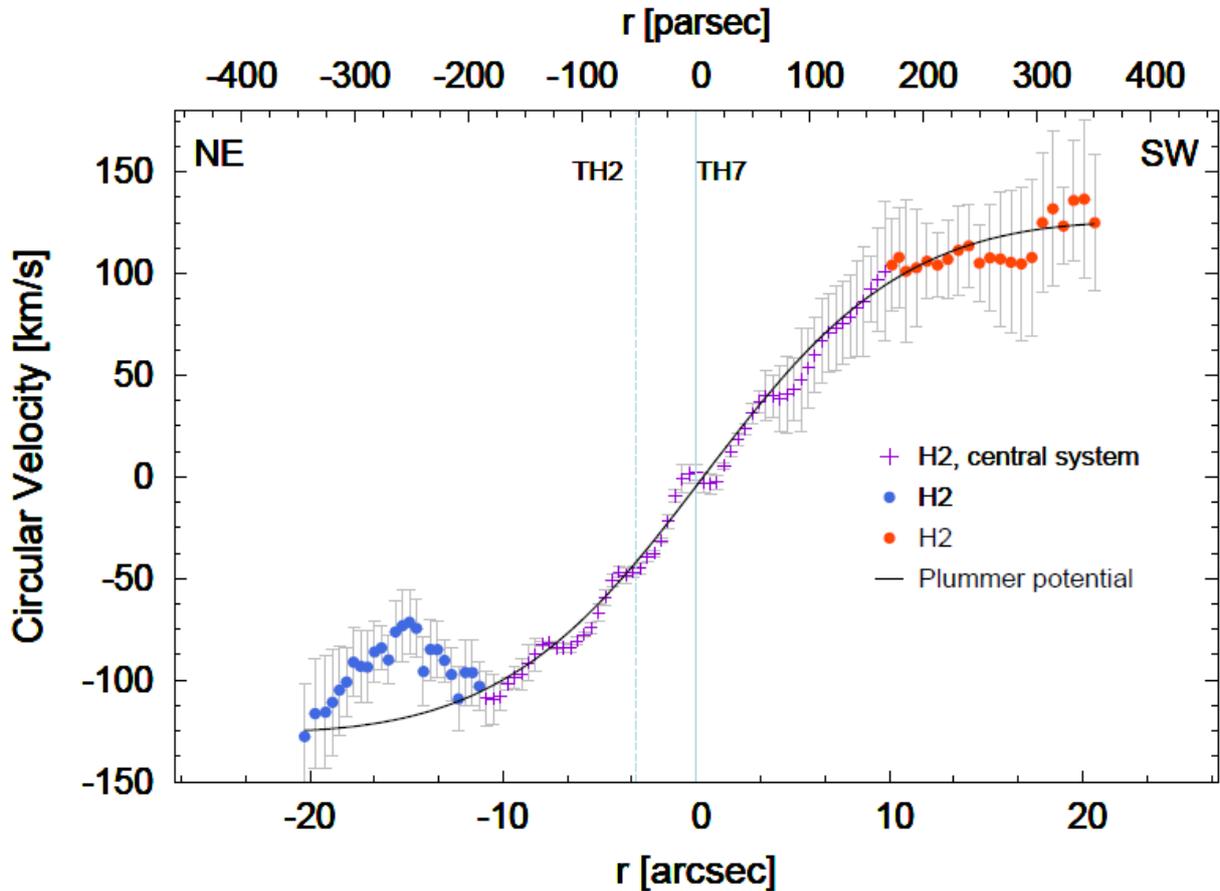

***Figure 17.*** $H_2$ *rotation curve along the major axis with the fitted rotation curve of a Plummer potential.*

The H$_2$ solid body rotation segment has a center of symmetry located one sampling element (0.36") toward the southwest. As the main component in the rotation curve is solid-bodylike, we fitted the rotation curve with a Plummer potential (Plummer 1911) shown in Figure 17. This model has the same center of symmetry as the solid body rotation segment. The best fit involves a scale radius of 0.25 Kpc and a total mass of 2.5x10$^9$(sin i)$^{-1}$ M$_\odot$. Considering that the circumnuclear disk is very near to an edge-on configuration (see Figure 8) and the high inclination (76°) of the NGC 253 disk (Hlavacek-Larrondo et al. 2011), then the inclination factor in the mass estimation is near to unity. The contribution of a global disk was found to be negligible at this circumnuclear scale.

The rotational component corresponding to the central Plummer potential was subtracted from the observed rotation curve in the innermost 5" (Figure 18). The residual velocity curve is also displayed at the bottom of the same figure where the S-shaped kinematic perturbation is evident and mimics a sinusoidal function. For reference, the positions of TH2 and the IRC have been marked in the figure. Note that the center of symmetry of S-shaped velocity perturbation is located one sampling element (0.18") toward the southwest of the IRC peak. As can be seen in Figure 18, the residual velocities mimic a sinusoidal function. Therefore, this perturbation was fitted with a sine function which is centered at 0.13" SW of the IRC. We considered the innermost resolved velocities as those corresponding to a radius of three spatial sampling elements and about one seeing diameter, i.e. at 0.54" from the center of symmetry of the rotation curve. If the symmetric residual is due to Keplerian motion, then its amplitude is consistent with a mass of (0.3±0.4)x10$^6$(sin i)$^{-1}$ M$_\odot$ within the resolved 9 pc radius.

Apart from the mentioned central S-shaped perturbation surrounding the IRC position, the rotation curve presents some other undulations with respect to the solid body rotation. The last ones are less pronounced than the central one. These kinematic perturbations are located in the following radial positions from IRC: 2.5"- 5.1" SW (± 6 Km/s), 2.9"- 5.0" NE (± 9 km/s), and 6.2"- 9.0" (± 6 km/s). The first one, located in the SW, is coincident with the edge of a dust lane. There are not any knots observed in NIR or MIR images at that position. The second one is consistent with the TH2 position (and green plume in Figure 4). This perturbation has almost half the amplitude of than that in the IRC position. Finally, the last one could be associated with A5 and A6 knots. Note that both perturbations close to the IRC are equidistant from it.

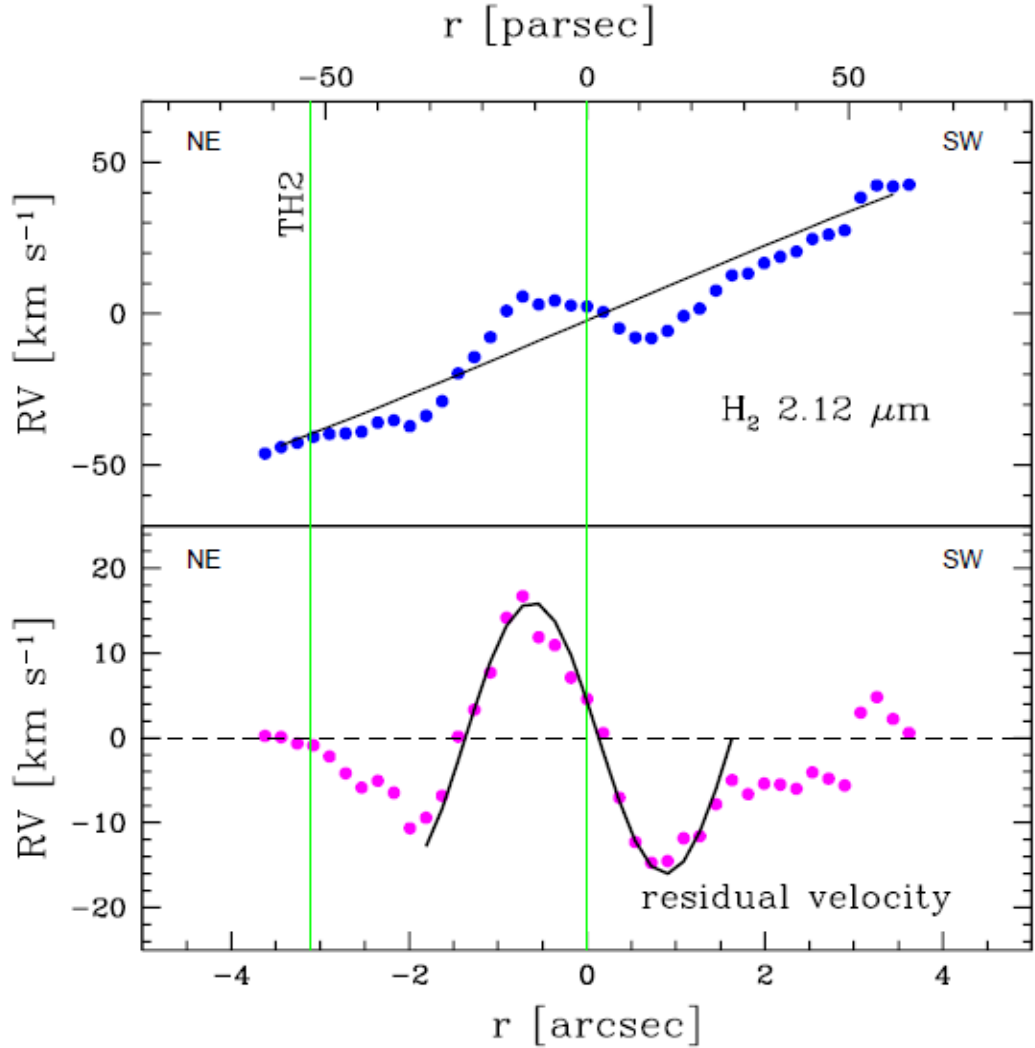

*Figure 18. Top: rotation curve of the nuclear region and global Plummer potential fitting. Bottom: residual velocity resulting from the subtraction between the rotation curve and Plummer potential fitting. A sinusoidal function has been fitted. For the circumnuclear region the signal to noise ration of the H$_2$ emission is always signal-to-noise ratio > 50; the average uncertainty was 4 kms$^{-1}$.*

## 4. Discussion

The nuclear region of NGC 253 presents a complex scenario in which the star formation processes, dust extinction, and galactic winds have obstructed the determination of the real nucleus of this galaxy. Therefore, the existence of a supermassive black hole in the nucleus of the nearest starburst galaxy is still not confirmed. Müller-Sánchez et al. (2010) determine the stellar kinematic center (SKC) in the NIR K band and describe in detail the nuclear region objects. They propose two potential nucleus candidates that could be associated with the SKC. One of them is the strongest radio source TH2 (Turner & Ho 1985). Considering that TH2 does not have any optical, IR, or X-ray counterpart, Fernández-Ontiveros et al. (2009) suggest that it harbors a Milky-Way-sized dormant black hole. A less considered nucleus candidate is the strongest hard X-ray source (X-1 in Müller-Sánchez et al. 2010), which has only been detected at high energies (> 2 keV). They suggest the possibility of X-1 being a hidden LLAGN, or maybe being associated with an NSC (corresponding to their IR-11 source with a possible radio counterpart TH6).

In this work, we present new evidence that points out to the IRC as a new candidate for the NGC 253 nucleus. Figure 19 depicts some important results. The stellar and molecular gas kinematical centers are plotted as well as the solid body rotation region. The bar component is drawn as a dotted line and crossing through the slit in the bar/disk symmetry center position. For reference, the TH2 and IRC positions are pointed out. The optical outflow galactic wind directions are indicated (Westmoquette et al. 2011). Additionally, the centers of the main structures, relative to the IRC position, are listed in Table 4.

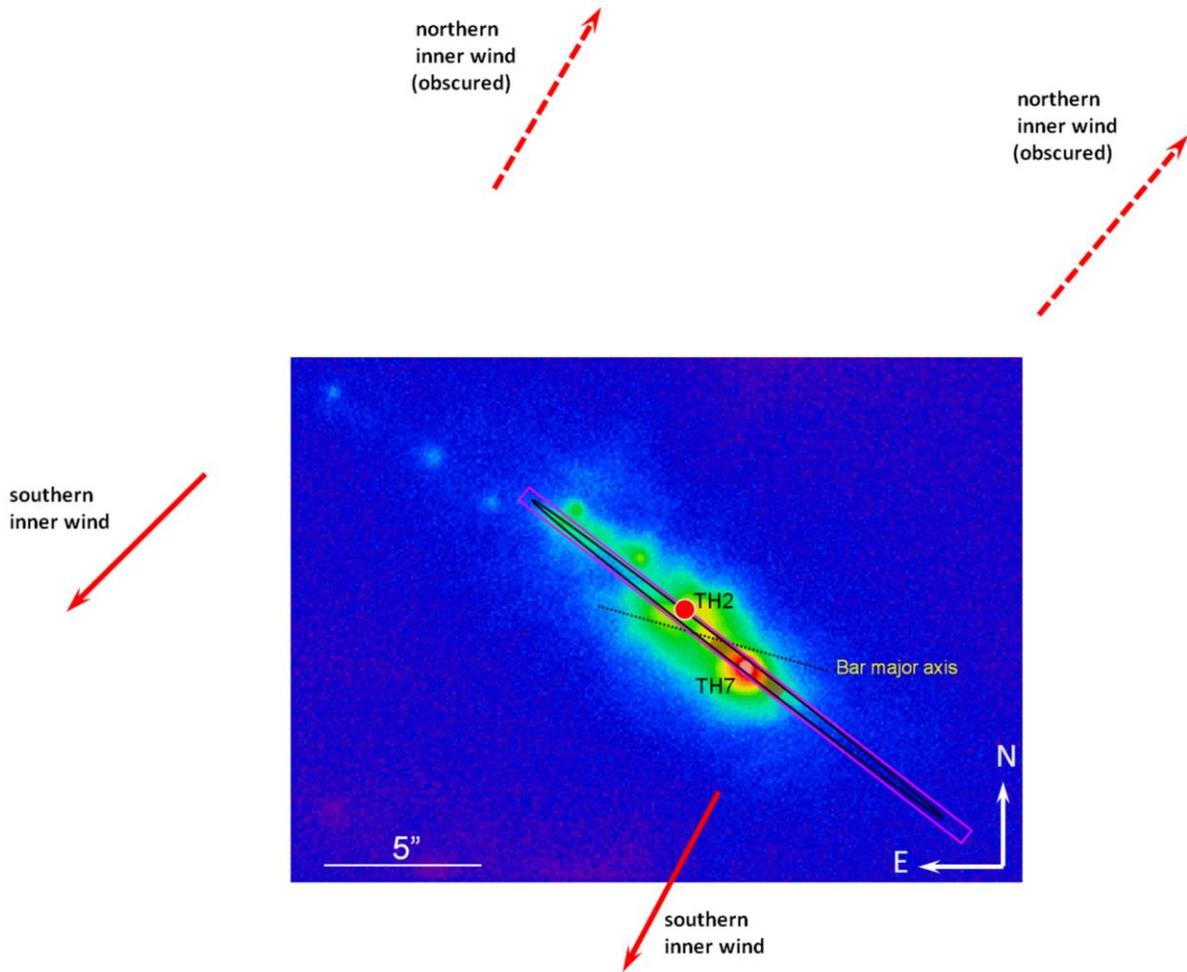

*Figure 19.* T-ReCS image of the nuclear region of the NGC 253, which resulted from adding the MIR images in the filters listed in Table 1. The spatial range that corresponds to the "S" shape in the radial velocity curve is indicated with a reddish semitransparent segment along the slit position angle and covering the TH7 position. Also marked are the rigid rotation zone (magenta rectangle) and the region associated with the profile of the narrow component stellar disk (black ellipse), with spatial dimensions reported in the text. The red arrows delimit the inner Hα wind outlined in Westmoquette et al. (2011).

**Table 4**
Offsets between the IRC and Diverse Structures

| Distance (″) | Symmetry Center | Method |
|---|---|---|
| 7.4 NE | $r^{1/4}$ law fitting of a bulge component | Continuum radial profile |
| 3.6 NE | Stellar broad component (Sérsic law with $n = 0.5$) | Continuum radial profile |
| 0.5 NE | Stellar exponential disk | Continuum radial profile |
| 0.36 SW | $H_2$ solid body rotation segment/Plummer potential model | $H_2$ radial velocities |
| 0.18 SW | Rotation curve of the nuclear region | $H_2$ radial velocities |
| 0.13 SW | Sinusoidal fit to nuclear residual velocity | $H_2$ radial velocities |
| 3.7 NE | Ks bulge component | NIR radial profile |
| 2.0 NE | Bar and global disk | NIR radial profile |
| 6.9 SW | Central color structure | NIR color profile |
| 2.6 NE | Stellar kinematic center | Muller-Sanchez et al. (2010) |
| 3.3 NE | Strongest radio source (TH2) | Turner & Ho (1985) |

**Note.** Col. (1): distance to the IRC, col. (2): symmetry center, col. (3): determination method.

First of all, the IRC is the strongest source in any observed MIR and NIR bands in the central 100 pc of the galaxy as well as the brightest soft X-ray source in the same region (0.5-10 keV, Müller-Sánchez et al. (2010)). Moreover, it is coincident with the bright radio source TH7 (Fernádez-Ontiveros et al. 2009). Kornei & McCrady (2009) established that IRC is the core of a massive star cluster of $1.4 \times 10^7$ $M_\odot$ with an estimated age of 5.7 Myr. They found a mixed stellar population suggesting a continuous star formation history or one punctuated by multiple bursts. This complexity in the stellar constituents plus the kinematic features that we present (the S-shaped radial velocity residual, the line profile asymmetries, and the ~ 200 km/s of FWHM), indicate a much more complex object than just a giant SSC.

From our kinematical analysis, the $H_2$ emission molecular gas rotation center is placed 0.18" SW of the IRC position. Namely, the molecular gas kinematic center, within the positional uncertainty of less than 0.2", can be linked with IRC rejecting its association with TH2 and the SKC (to 3.5" and 2.8" respectively; see Figure 19). Besides reinforcing the predominance of the IRC neighborhood, the main $H_2$ rotation curve residual is located 0.13" SW of the IRC position (Figure 19). The radial velocity residual has a sinusoidal shape with an amplitude of 16 kms$^{-1}$ (Figure 18). The estimated pointlike mass within 9 pc at that position is $0.3 \times 10^6$ $M_\odot$. In addition to the preferential location of IRC at the rotation center of the molecular gas disk, the continuum spatial profile of the spectra shows the IRC surrounded by a compact massive stellar disk (§3.3, Figure 19). However, there is another stellar structure in the nuclear region (less luminous and more diffuse than the disk component) whose symmetry center is at 0.3" from the TH2 position.

The slit goes across quite close to the TH2 position (it is located at the edge of the slit), which is enough to detect a kinematic perturbation produced by the presence of a massive object at that position (Figure 1). In Figure 17, a small undulation in the rotation curve is observed near the TH2 position. The amplitude of the observed kinematic perturbation is (9 ± 5) km/s, considering the spatial resolution of 0.5", the largest detected pointlike mass in the TH2 position would be $2 \times 10^5$ $M_\odot$. Moreover, in accordance with the results of Fernández-Ontiveros et al. (2009), we did not detect any compact MIR source on the TH2 spot down to a flux of 40 mJy (Qa band) in any of the four T-ReCS bands. The point source detection threshold was determined as an average of the measured fluxes of the two faintest pointlike objects (A7 and A9). However, there are several fainter diffuse sources detected in the circumnuclear region, with fluxes fainter than 10 mJy (Table 2). On the other hand, the slit marginally touches the SKC position, but in this case, no kinematic perturbations are detected at that radius resulting in a lower mass limit for a potential presence of a massive object in the SKC position.

From our MIR imaging analysis, some distinctive structures are observed near the IRC. In the T-ReCS [NeII] 12.8 μm emission, an arc-shell structure is clearly perceived surrounding the IRC (Figure 5). Moreover, some plumelike features are distinguished emerging from the IRC toward SW in the [NeII] 12.8 μm and Si 8.8 μm T-ReCS images (Figure 4) and toward the south in the Paα HST image (Figure 1). All of these features are consistent with an outflow process taking place in the IRC neighborhood. It is worth noting that some peculiar plumelike features are observed in [NeII] 12.8 μm emission near the TH2 position with a N-S orientation (Figure 4).

From our NIR imaging analysis, we found that the bar and global disk symmetry center is at 2" NE of the IRC, between the TH2 and IRC nucleus candidates (Figure 19). However, the Ks-band bulge component is off-centered by 3.7" NE of the IRC; in this case, the radio source TH2 is the object nearest to the bulge center (Figure 9). In the color maps of the central region of Figure 8, one can see a color disk more extended toward the SW side of the IRC where the dust filaments are evidently emerging from it in contrast with the central disk observed in the NIR band images (Figure 7), which is more extended toward the NE of the IRC. This is consistent with the analysis of the Ks and Y-Ks profiles along the bar major axis (Figure 9). The central color structure is clearly prolonged in the SW direction (centered at 7" SW of the IRC) while the Ks bulge component is compacted in the NE direction (centered at 3.7" NE of the IRC). Furthermore, practically we cannot find bright knots in the SW side of the IRC. These results can be explained if the SW side of the IRC is highly extinguished. In that case, the symmetry center of the Ks-band bulge component would be artificially shifted toward the NE direction, explaining the difference reported between the bar/disk and bulge symmetry centers.

Finally, we present strong evidence pointing to the IRC as the main outflow source in the nuclear region, which feeds the well-known circumnuclear outflow (Weaver et al. (2002); Westmoquette et al. 2011; Bolatto et al. 2013; Monje et al. 2014). The kinematical perturbation observed at the IRC position (Figure 18) could be due to the presence of outflowing gas from the IRC. The maximum outflow velocity is located about 14 pc from the IRC, which is consistent with the radius of a shell observed around the IRC at [NeII] 12.8 μm emission using T-ReCS at Gemini (Figure 5). It is important to mention that a tri-axial spherical component could generate an S-shaped rotation curve depending on the orientation of the spherical component (Arnold et al. 1994; Coccato et al. 2004). Moreover, we cannot discount the possibility that the feature may be associated with local rotation around a compact object at the IRC position. It is worth noting that a close inspection of the published stellar radial velocity field (Müller-Sánchez et al. 2010) shows a pair of positive and negative lobes at the position of the A1 complex, centered on the IRC with an uncertainty of 0.2", derived from the coordinate scales in the published map and the uncertainty of the IRC location in our data. This would be consistent with the rotation of the stellar disk detected at the continuum spectrum profile. Additionally, Westmoquette et al. (2011) reported a blueshifted broad component in the Hα emission line from the VMOS field which includes the nuclear region of NGC 253. We also observe this blueshifted asymmetry in the Brγ line profile. Moreover, our kinematical data allowed us to constrain the detection of the blueshifted component only in the IRC surroundings, being more emphasized toward the IRC position where it reaches the maximum intensity. Monje et al. (2014) found a P-Cygni profile in the hydrogen fluoride (HF) spectrum in the nuclear region of NGC 253 using the Heterodyne Instrument for the Far-Infrared on board the *Herschel Space Observatory*. They argue an outflow origin for the P-Cygni feature. However, due to the low spatial resolution (17" beam size), they cannot point out the position where the outflow is originating. Completing the outflow picture, we present the 2D spectrum zoomed in the Brγ emission line (Figure 20).

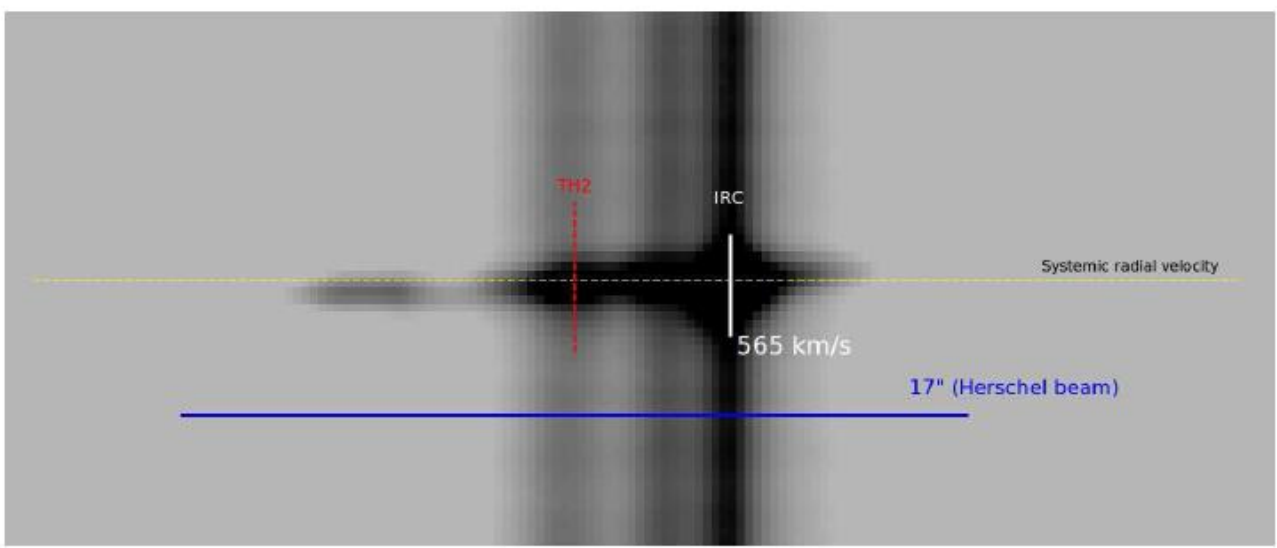

*Figure 20. Zoom of the 2D spectrum, showing in detail the Brγ emission line. The asymmetry in the line profile is evident when compared with the galaxy systemic radial velocity, which is shown in the spectrum. Also is plotted the radial velocity width of 565 km/s, considering a line intensity up to 10 % of the peak intensity value. The Herschel beam size (17") is superimposed in the plot for comparison.*

This line exhibits a clear signal of mass ejection from the IRC surroundings, estimating the velocity amplitude of 565 kms$^{-1}$ (considering the line intensity up to 10 % of the peak intensity value). For reference, the TH2 position and systemic velocity are marked. In that figure, the *Herschel* HF beam size is plotted. It is evident that the IRC is the most probable source responsible for the HF P-Cygni profile of the nuclear outflow.

Considering the possibility of the IRC as being the nucleus of NGC 253, we wonder if IRC could harbor a supermassive black hole. Fernández-Ontiveros et al. (2009) built SED from optical to MIR of several knots in the nuclear regions, including the knot corresponding to the IRC (knot 4 in their work). They noticed the similarity of the SEDs for all knots and found that the SEDs were well reproduced by considering an important contribution of very young stellar objects. From this work, the SED of the IRC does not seem to evidence an AGN contribution. In order to analyze the emission nature of the IRC, we took into account emission line flux ratios, which may distinguish between the starburst or AGN emission. Particularly, we employed the (log($H_2$/Brγ) and log(FeII/Paβ)) proposed by Riffel et al. (2010) and Larkin et al. (1998). Kornei & McCrady (2009) listed the line fluxes of the region surrounding the IRC which allowed us to place them in the mentioned NIR diagnostic diagram: log(FeII/β) = -0.41, log($H_2$/Brγ) = -0.76 (due to the spectral range of our spectra, we only have data for the last ratio, which in our case is (-0.87±0.09), in accordance with Kornei & McCrady's (2009) ratio). This value undoubtedly places the knot containing the IRC in the starburst region of the diagram. Moreover, in our spectrum we find no emission line features at the Ks band that could unarguably be associated with a hidden low luminosity AGN. Therefore, there are no clues that suggest the presence of a bright AGN at the IRC position. The presence in the NIR spectrum of the CO band at the edge of the observed spectral range, and of the Brγ and He I line (Figure 10) imply the co-existence of stars with ages of more than a few $10^7$ years, together with young stars of a few million years old, in agreement with what has been stated by Kornei & McCrady (2009) who found a mixed stellar population in the central cluster and conclude that the star formation is happening recursively at the IRC region. As a consequence, the star formation enhancement could be masking the presence of a low luminosity AGN. Besides, the Brγ line width at the IRC (FWHM ~ 200km/s, Figure 20) would be considered typical for the narrow lines in a low luminosity AGN should the emission be observed at higher luminosity in a farther galactic nucleus. Therefore we cannot rule out the hypothesis that the IRC core is harboring a very low luminosity AGN. In §3 we reported the innermost resolved mass, which for the 0.5" seeing only encompasses the core of the cluster and serves as an indication of the maximum order of magnitude of the hypothetical black hole mass. If we consider the radial velocity difference at a 2" radius (~ 27km/s), a radius in which the radial velocity curve beam smearing should be negligible and the whole cluster mass should be detected, then the $H_2$ rotation curve gradient yields an estimated Keplerian inner mass of $0.5 \times 10^7$ (sin i)$^{-1}$ $M_\odot$, which is within the same order of magnitude as the mass estimated from Kornei & McCrady (2009). Their mass estimation comes from spectroscopy with a spatial resolution of 1.2". A slight overestimation of the extinction values in Kornei & McCrady (2009), or an inner disk oriented 40° with respect to the plane of the sky would account for the estimated mass difference. The innermost mass estimation is quite low and we could just be detecting the unresolved core of the massive stellar cluster. However, the presence of a $1 \times 10^6$ $M_\odot$ SMBH is not excluded within a $1 \times 10^7$ $M_\odot$ cluster. Moreover, it would be an eventual by-product of the dynamical evolution of such a massive nuclear star cluster.

All the results listed above should be enough to consider the IRC as a solid candidate for the NGC 253 nucleus. Although some galaxies would show evidence of two coexisting galactic nuclei (e.g. M31, Kormendi & Bender (1999); M83, Mast et al. (2006)) we think this is not the case for NGC 253. The described scenario is not favorable to the possibility of a double nucleus that originated in a minor merger event. Mainly, the substantial amount of gas supplied to the nuclear region (M($H_2$)$_{out}$ =$1 \times 10^7$ $M_\odot$, dM/dt=6.4 $M_\odot$ yr$^{-1}$, Monje et al. (2014); dM$_{mol}$/dt = 9 $M_\odot$ yr$^{-1}$, Bolatto et al.

2013) should be enough to feed any massive object living in that region and make it easier to detect as is the case with the IRC. However, no counterpart was observed for the radio source TH2. In the case of the X-ray source X-1, which has been mentioned as possible low luminosity AGN (Müller-Sánchez et al. 2010), it does not seem to be likely that it is a supermassive object. We do not detect any distinguished spectroscopic feature (in emission line ratios and kinematics) at the region of X-1, and the NIR and MIR brightnesses are not conspicuous in comparison with other sources in the field.

## 5. Concluding Remarks

Until now, the IRC has been associated with an SSC (Kornei & McCrady 2009). Beyond the SSC mixed stellar population, the results obtained in the present work suggest a much more complex object than just a giant SSC.

The IRC is the brightest IR (NIR and MIR) source and the most powerful soft X-ray emission source. Surprisingly, there are no pointlike sources or kinematic features associated with the brightest radio source TH2. It is important to note that the radio source associated with the IRC (TH7) is the second strongest radio source.

From the kinematic data furnished by F2, the H2 emission rotation curve has been obtained with suitable resolution. That curve is well represented by a solid body rotation curve whose center is coincident with the IRC position. The most important feature of the kinematic studies is the presence of a sinusoidal kinematic perturbation around the IRC. The similar feature observed at the TH2 position is substantially less prominent. Not only the $H_2$ molecular disk is centered at the IRC position, but also the nuclear stellar disk component surrounds the IRC.
We also found that the IRC region shows the highest star formation rate and the presence of radial velocity turbulences. The observed asymmetry of the line profile suggests that an outflow process is taking place in the IRC surroundings. The detection by Bolatto et al. (2013) of a strong outflowing gaseous component involving the whole central galaxy region favors the outflow nature of the S-shaped feature in the H2 rotation curve. This is reinforced by the thermalization found in the IRC location and the shell-like structures detected in the MIR images (at an average spatial resolution of 5 pc). These shell-like structures extend in the radial range of 11-16 pc from the IRC, which therefore shows the most conspicuous outflow characteristics in the NGC 253 nuclear region, both in kinematics and morphology.

It has been so far unclear which is the true dynamical nucleus of the galaxy to the point that there is no strong evidence that the galaxy harbors a supermassive black hole co-evolving with the starburst. In the present work kinematic, spectrophotometric, and morphological evidence has been shown which supports the hypothesis of IRC being the galactic nucleus, the most massive object, and the major source of the nuclear starburst outflow. The off-center position of the IRC and the circumnuclear disk with respect to the bulge imply a decoupling of the central gas and nuclear cluster from the large-scale structures, as predicted by Emsellem et al. (2015) for a gas-rich barred galaxy modeled at sub-parsec resolution. These authors show that the lack of symmetry in the mass distribution triggers the formation of gas clumps which form stars or get disrupted, providing a mechanism for angular momentum removal. Supernovae shells and filaments also contribute in the angular momentum removal of the gas and the feeding of the massive central object. Moreover, the surrounding structure of the IRC is massive enough to harbor about million solar masses within a radius of a few parsecs, which eventually could evolve to a supermassive black hole if that is not yet the case.

We gratefully acknowledge the anonymous referee for useful suggestions, which helped to improve the presentation of this study. We thank Bernadette Rodgers for setting the T-ReCS Legacy program GS-2011B-Q-84, and Horacio Dottori and Facundo Albacete-Colombo for fruitful discussions on the NGC 253 scenario. We acknowledge grant support from CONICET (PIP 0523), ANPCyT (PICT 835), and SeCyT-UNC (05/N030). This research was based on observations obtained at the Gemini Observatory, which is operated by the Association of Universities for Research in Astronomy, Inc., under a cooperative agreement with the NSF on behalf of the Gemini partnership: the National Science Foundation (United States), the Science and Technology Facilities Council (United Kingdom), the National Research Council (Canada), CONICYT (Chile), the Australian Research Council (Australia), Ministério da Ciência, Tecnologia e Inovação (Brazil), and Ministerio de Ciencia, Tecnología e Innovación Productiva (Argentina). We acknowledge grant support from CONICET (PIP 0523), ANPCyT (PICT 835) and SeCyT-UNC.

## REFERENCES

Alonso-Herrero, A., Rieke, G. H., Rieke, M. J. & Kelly, D. M. 2003, AJ, 125, 1210

Arnold, R., de Zeeuw, P. T. & Hunter, C. 1994, MNRAS, 271, 924

Binney, J., & Merrifield, S., 1998, Galactic Astronomy, Princeton University Press

Black, J. H., & van Dishoeck, E. J. 2006, ApJ, 322, 412

Bolatto, A. D., Warren, S. R., Leroy, A. K., Walter, F., Veilleux, S., Ostriker, E. C., Jurgen, O., Zwaan, M., Fisher, D. B., Weiss, A., Rosolowsky, E. & Hodge, J. 2013, Nature, 499, 450

Chen, Yan-Mei, Wang, Jian-Min, Yan, Chang-Shou, Hu, Chen & Zhang, Shu 2009, ApJ, 695, L130

Cid-Fernandes, R., Heckman, T., Schmitt, H., Gonzalez Delgado, R. M. & Storchi-Bergmann, T. 2001, ApJ, 558, 81

Coccato, L., Corsini, E. M., Pizzella, A., Morelli, L., Funes, J. G. & Bertola, F. 2004, A&A, 416, 507

Davies, R. L., Rich, J. A., Kewley, L. J. & Dopita, M. A. 2014, MNRAS, 439, 3835

de Vaucouleurs, G. 1959, Handbuch der Physik, 53, 275

Díaz, R.J., Dottori, H., Agüero, M.P., Mediavilla, E., Rodrigues, I. & Mast, D. 2006, ApJ, 652, 1122

Díaz, R.J., Gomez, P., Schirmer, M., Navarrete, F., Stephens, A., Bosch, G., Gaspar, G., Camperi, J. A. & Günthardt, G. 2013, BAAA, 56, 457

Eikenberry S., et al., 2008, Ground-based and Airborne Instrumentation for Astronomy II. Edited by McLean, Ian S.; Casali, Mark M. Proceedings of the SPIE, Volume 7014, article id. 70140V

Emsellem, E., Renaud, F., Bournaud, F., Elmegreen, B., Combes, F. & Gabor, J. M. 2015, MNRAS, 446, 2468

Engelbracht, C. W., Rieke, M. J., Rieke, G. H., Kelly, D. M. & Achtermann, J. M. 1998, ApJ, 505, 639

Erben, T., Schirmer, M., Dietrich, J. P., Cordes, O., Haberzettl, L., Hetterscheidt, M., Hildebrandt, H., Schmithuesen, O., Schneider, P., Simon, P., Deul, E., Hook, R. N., Kaiser, N., Radovich, M., Benoist, C., Nonino, M., Olsen, L. F., Prandoni, I., Wichmann, R., Zaggia, S., Bomans, D., Dettmar, R. J. & Miralles, J. M. 2005, Astronomische Nachrichten, 326, 432

Falcón-Barroso, J., Ramos Almeida, C, Boker T., Schinnerer, E., Knapen, J. H., Lancon, A. & Ryder, S. 2014, MNRAS, 438, 329

Fernández-Ontiveros, J. A, Prieto, M. A. & Acosta-Pulido, J. A. 2009, MNRAS, 392, L16

Gómez P. L., Díaz, R. J., Pessev P., et al., 2012, American Astronomical Society, AAS Meeting 219, 413.07

Gonzalez Delgado, R. M., Munoz Marin, V. M., Perez, E., Schmitt, H. R. & Cid Fernandes, R. 2009, Ap&SS, 320, 61

Hlavacek-Larrondo, J., Marcelin, M., Epinat, B., Carignan, C., de Denus-Baillargeon, M.-M., Daigle, O. & Hernandez, O. 2011, MNRAS, 416, 509

Hollenbach, D. & McKee, C. F. 1989, ApJ, 342, 306

Iodice, E., Arnaboldi, M., Rejkuba, M, Neeser, M. J., Greggio, L., Gonzalez, O. A., Irwin, M. & Emerson, J. P. 2014, A&A, 567, 86

Johnson, K. E. 2005, in IAU Symposium, Vol. 227, Massive Star Birth: A Crossroads of Astrophysics, ed. R. Cesaroni, M. Felli, E. Churchwell & M. Walmsley, 413–422

Karachentsev, I. D., Grebel, E. K., Sharina, M. E., Dolphin, A. E., Geisler, D., Guhathakurta, P., Hodge, P. W., Karachentseva, V. E., Sarajedini, A. & Seitzer, P. 2003, A&A, 404, 93

Kennicutt, Jr., R. C. 1998, ARA&A, 36, 189

Kormendi, J. & McCrady, R. 1999, ApJ, 522, 772

Kornei, K. A. & McCrady, N. 2009, ApJ, 697, 1180

Larkin, J. E., Armus, L., Knop, R. A., Soifer, B. T. & Matthews, K. 1998, ApJS, 114, 59

Lenc, E. & Tingay, S. J. 2006, AJ, 132, 1333

Levenson, N. A., Weaver, K. A. & Heckman, T. M. 2001, ApJ, 550, 230

Maloney, P. R., Hollenbach, D. J. & Tielens, A. G. G. M. 1996, ApJ, 466, 561

Mast, D., Díaz, R. J. & Agüero, M. P. 2006, AJ, 131, 1394

Mentuch, E., Abraham, R. G., Glazebrook, K., McCarthy, P. J., Yan, H., O'Donnell, D. V., Le Borgne, D., Savaglio, S., Crampton, D., Murowinski, R., Juneau, S., Carlberg, R. G., Jørgensen, I., Roth, K., Chen, H.-W. & Marzke, R. O. 2009, ApJ, 706, 1020

Monje, R. R., Lord, S., Falgarone, E., Lis, D. C., Neufeld, D. A., Phillips, T. G. & Güsten, R. 2014, ApJ, 785, 22

Mouri, H. 1994, ApJ, 427, 777

Müller-Sánchez, F., González-Martin, O., Fernández-Ontiveros, J. A., Acosta-Pulido, J. A. & Prieto, M. A. 2010, ApJ, 716, 1166

Norman, C. & Scoville, N. 1988, , 332, 124

Plummer, H. C. 1911, MNRAS, 71, 460

Rafanelli, P., La Mura, G., Bindoni, D., Ciroi, S., Cracco, V., Di Mille, F. & Vaona, L. 2011, Baltic Astronomy, 20, 419

Reunanen, J., Kotilainen, J. K. & Prieto, M. A. 2002, MNRAS, 331, 154

Riffel, R. A., Storchi-Bergmann, T. & Nagar, N. M. 2010, MNRAS, 404, 166

Rodrigues, I., Dottori, H., Díaz, R.J., Agüero, M.P. & Mast, D. 2009, AJ, 137, 4083

Rodríguez-Ardila, A., Pastoriza, M. G., Viegas, S., Sigut, T. A. A. & Pradhan, A. K. 2004, A&A, 425, 457

Rodríguez-Ardila A., Riffel, R. & Pastoriza, M. G. 2005, MNRAS, 364, 1041

Rodríguez-Ardila, A., Riffel, R., Pastoriza, M. G., Maraston, C. & Carvalho, E. A. 2009, in The Starburst-AGN Connection. ASP Conferences Series, Vol. 408, 203

Schirmer, M. 2013, ApJS, 209, 21

Sternberg, A. & Dalgarno, A. 1989, ApJ, 338, 197

Sugai, H., Davies, R. I. & Ward, M. J. 2003, ApJ, 584, 9

Telesco, C. M., Pina, R. K., Hanna, K. T., Julian, J. A., Hon, D. B. & Kisko, T. M. 1998. SPIE 3354, 534

Turner, J. L. & Ho, P. T. P. 1985, ApJ, 299, L77

Ulvestad, J. S., & Antonucci, R. R. J. 1997, ApJ, 488, 621

Watson, A. M., Gallagher, III, J. S., Holtzman, J. A., Hester, J. J., Mould, J. R., Ballester, G. E., Burrows, C. J., Casertano, S., Clarke, J. T., Crisp, D., Evans, R., Griffiths, R. E., Hoessel, J. G., Scowen, P. A., Stapelfeldt, K. R., Trauger, J. T. & Westphal, J. A. 1996, AJ, 112, 534

Weaver, K. A., Heckman, T. M., Strickland, D. K. & Dahlem, M. 2002, ApJ, 576, L19

Westmoquette, M. S., Smith, L. J. & Gallagher, J. S. III 2011, MNRAS, 414, 3719